\documentclass[12pt]{article}
\usepackage{amsfonts}
\usepackage{mathrsfs}
\usepackage{amsthm}
\usepackage{multirow}
\usepackage{bbding}
\usepackage{amssymb}
\usepackage{amsmath}
\usepackage{graphicx,color}
\usepackage{rotating}
\usepackage[justification=justified,singlelinecheck=off]{caption}
\usepackage{bm}

\DeclareMathOperator*{\argmin}{arg\,min}
\newcommand\bbf[1]{\mathchoice{\hbox{\boldmath$\displaystyle#1$}}
{\hbox{\boldmath$\textstyle#1$}} {\hbox{\boldmath$\scriptstyle#1$}}
{\hbox{\boldmath$\scriptscriptstyle#1$}} }
\newcommand{\bea}{\begin{eqnarray}}
\newcommand{\eea}{\end{eqnarray}}
\newcommand{\Bea}{\begin{eqnarray*}}
\newcommand{\Eea}{\end{eqnarray*}}
\newcommand{\ba}{\begin{array}}
\newcommand{\ea}{\end{array}}
\newcommand{\bt}{\begin{tabular}}
\newcommand{\et}{\end{tabular}}
\newcommand{\btb}{\begin{table}}
\newcommand{\etb}{\end{table}}
\newcommand{\bc}{\begin{center}}
\newcommand{\ec}{\end{center}}
\newcommand{\beq}{\begin{equation}}
\newcommand{\eeq}{\end{equation}}

\makeatletter

\newcommand{\Rmnum}[1]{\expandafter\@slowromancap\romannumeral #1@}
\makeatother

\begin{document}

\title{Group-Average and Convex Clustering for Partially Heterogeneous Linear Regression
}
\author{Lu Lin$^1$, Jun Lu$^1$
and Chen Lin$^2$\footnote{The corresponding
author. Email: 13969154756@163.com. The research was
supported by NNSF projects (11571204 and 11231005) of China.}
\\
$^1$Zhongtai Securities Institute for Financial Studies\\ Shandong University, Jinan, China\\
$^2$School of Economics, Renmin University of China, Beijing, China}
\date{}
\maketitle

\vspace{-0.3cm}

\begin{abstract} \baselineskip=15pt

In this paper, a subgroup least squares and a convex clustering are introduced for inferring a partially heterogenous linear regression that has potential application in the areas of precision marketing and precision medicine. The homogenous parameter and the subgroup-average of the heterogenous parameters can be consistently estimated by the subgroup least squares, without need of the sparsity assumption on the heterogenous parameters. The heterogenous parameters can be consistently clustered via the convex clustering. Unlike the existing methods for regression clustering, our clustering procedure is a standard mean clustering, although the model under study is a type of regression, and the corresponding algorithm only involves low dimensional parameters. Thus, it is simple and stable even if the sample size is large. The advantage of the method is further illustrated via simulation studies and the analysis of car sales data.

\

{\it Key words:} Regression, heterogeneity, group-average, estimation consistency, convex clustering.



\end{abstract}

\newpage
\baselineskip=19pt

\setcounter{equation}{0}
\section{Introduction}
The theory and methodology of precision marketing focus mainly on
the problem of choosing the right strategic decision-making policies for selling the right
products to the right customers at the right time, such that the companies can increase their profits (see, e.g., Zabin et al.; 2004, and You et al.; 2015).
In the procedure of precision marketing through certain pathways such as the internet, the individual information of potential customers is usually employed to promote personalized products. To describe such a procedure by a linear regression, for example, we use $Y$ to denote the logarithm probability of purchasing a product by a customer, and use $Z$ to denote some characteristics of customers, such as product packaging and appearance, and customer's age and gender. It is worth pointing out that different characteristic levels may have
different effects on the consumer behavior. Thus,
besides the homogenous components, the related heterogenous components should be included in the regression to precisely characterize the individual effects.
We thus are interested in the following partially heterogenous linear regression model:
\begin{eqnarray}\label{(1.1)}Y_i =X_i^T\beta+Z_i^T\theta_i+\epsilon_i, \ i=1,\cdots, n.\end{eqnarray}
Here $Y_i,i=1,\cdots,n$, are the observations of $Y$, $(X_i,Z_i),i=1,\cdots,n$, are independent and identically distributed observations of $(X,Z)$ with $X$ and $Z$ being $d_X$-dimensional and $d_Z$-dimensional covariates respectively, and the errors $\epsilon_i,i=1,\cdots, n$, are mutually independent with $E[\epsilon_i|X_i,Z_i]=0$ and $\sigma^2=Var[\epsilon_i|X_i,Z_i]$. A notable feature is that there are both homogenous coefficient $\beta$ and heterogenous coefficients $\theta_i$ in the model for respectively valuating the common and individual effects. In regression analysis for the partially heterogenous linear regression model, the main goal is to consistently estimate the homogenous parameter $\beta$, and divide the heterogenous coefficients $\theta_1,\cdots,\theta_n$ into several groups such that in the same group, the corresponding parameters $\theta_i$ have the same value. Consequently, the specific treatments, such as individualized marketing strategies can be implemented for each subgroup.
This is just like the market segmentation the precision marketing relies on: a technique for breaking the market down into smaller, more specific blocks of customers with unique needs. Market segments can be very broad; women, for example, or they can be very specific; unmarried women over 50 with adopted children.
These imply the sparsity assumption on the set $\{\theta_1,\cdots,\theta_n\}$. Here the sparsity means that $\theta_1,\cdots,\theta_n$ can be separated into several subgroups such that in each subgroup the parameters $\theta_i$ have the same value (i.e., each subgroup represents a special customer group), and the number of the subgroups is small.


A commonly used approach to dealing with the heterogeneity is based on finite mixture models; for the related references on mixture models see, e.g., Everitt and Hand (1981), Banfield
and Raftery (1993), Hastie and Tibshirani (1996) and McNicholas (2010). Recently, Shen and He (2015) used the approach to subgroup analysis, and
You et al. (2015) provided an overview of the classical methodologies for customer classification in precision marketing, and presented a novel decision-making framework
for precision marking using data-mining techniques. The classical mixture model approach requires knowing the information of subgroups and a parametric assumption on the model.

In our model (\ref{(1.1)}), however, the grouping information of individual customers is unknown in advance. A popularly used way that does not need the grouping information is based on penalty or regularization, including the fused lasso (Tibshirani and Saunders; 2005) and the related versions (Bondell and Reich; 2008, Shen and Huang; 2010, Ke et al.; 2015, Guo et al.; 2010, and Guo et al.; 2016), and the convex clustering (Chi and Lange; 2014 and Boyd et al.; 2011).
Motivated by penalty-based methods, Ma and Huang (2016a) proposed a concave pairwise fusion methodology to estimate the homogenous parameter $\beta$ and group the heterogenous components $\theta_i$ for the special case of $Z_i\equiv 1$ for $i=1,\cdots,n$. Furthermore, Ma and Huang (2016b) extended the method to general cases. Such a methodology is a successful application of fused Lasso to the area of regression clustering. However, the related algorithm includes ``descent" and ``fusion" steps. It is complicated because in each descent step, the algorithm involves calculating the iterative solutions of $(\theta_1^T,\cdots,\theta_n^T)^T$ simultaneously, which actually is a penalized least square estimator of $(\theta_1^T,\cdots,\theta_n^T)^T$ with the dimension of $ n\times d_Z$. When the sample size $n$ is large, consequently, the dimension of $(\theta_1^T,\cdots,\theta_n^T)^T$ is high and thus the algorithm is not stable.
Another penalty-based method was introduced by Lin et al. (2017), who employed a two-step method to identify both the homogenous parameter $\beta$ and the maximum-risk subgroup of heterogeneous parameters $\theta_i$ and then to build an upper expectation regression. Such a method only can recognize the maximum-risk subgroup, without the result about identifying the other subgroups. On the other hand, all the aforementioned methods require the sparsity assumption on the set $\{\theta_1,\cdots,\theta_n\}$.

In the present paper, we introduce a subgroup least squares to consistently estimate the homogenous parameter $\beta$ and the subgroup-averages of the heterogenous parameters $\theta_i$. Based on the consistent estimators, we suggest a convex clustering to identify the subgroups of $\theta_i$. Our methodology has the following salient features:
\begin{itemize}\item Unlike the algorithms of the existing methods for regression clustering (see, e.g., Ma and Huang; 2016a and 2016b), our algorithm is a standard convex clustering, without involving regression clustering. Then, the algorithm procedure only depends on the iterations of $d_Z$-dimensional parameters in each step. Hence, it is simple and stable even the sample size $n$ is large.
\item On the other hand, the consistency of the estimator of $\beta$ is free of the sparsity assumption on the set $\{\theta_1,\cdots,\theta_n\}$. It makes sense for the case when our goal is only to estimate the homogenous effect $X^T\beta$. However, the aforementioned methods cannot achieve this goal if without the sparsity assumption.
\item Moreover, such a method can be extended to the case of estimating equation-based model.
\end{itemize}

The remainder of the paper is organized in the following way. In Section 2, the subgroup least squares is first defined, and then the consistent estimators of $\beta$ and the subgroup-average of $\theta_i$ are constructed, and finally the classification for the heterogenous parameters $\theta_i$ and the corresponding algorithm are introduced. The theoretical properties of the proposed approach are given in Section 3. In Section 4, the finite sample properties of the proposed procedures are evaluated via simulation studies, and the proposed methods are further illustrated by analyzing car sales data. The proofs of the theorems and lemmas are provided in the appendix.

\setcounter{equation}{0}
\section{Methodology}

\subsection{Parameter estimation}

Let $\mathscr A$ be an arbitrary subgroup of the index set $\mathscr G=\{1,\cdots,n\}$, and $|\mathscr A|$ be the size of $\mathscr A$, i.e., the number of elements of $\mathscr A$. Denote by $\overline W_{\mathscr A}$ the subgroup-average of vectors or matrices $W_i$ confined on $\mathscr A$, namely, $$\overline W_{\mathscr A}=\frac{1}{|\mathscr A|}\sum_{i\in \mathscr A} W_i.$$ For example, the subgroup-average of $\theta_1,\cdots,\theta_n$ confined on $\mathscr A$ is denoted by $\overline \theta_{\mathscr A}=\frac{1}{|\mathscr A|}\sum_{i\in \mathscr A} \theta_i$. Particularly, when $\mathscr A=\mathscr G$, $\overline W_{\mathscr G}$ is written as $\overline W$.

For model (\ref{(1.1)}), the subgroup least squares objective function confined on $\mathscr A$ is defined by
$$\frac{1}{|\mathscr A|}\sum_{i\in \mathscr A}\left(Y_i-X_i^T\beta-Z_i^T\theta_i\right)^2.$$ By taking the derivatives of the above objective function with respect to $\beta$ and $\theta_i$ for $i\in \mathscr A$, and summing them over $i\in \mathscr A$, we get the following estimating equations:
\begin{eqnarray}\label{(2.1)}(\overline{XY})_{\mathscr A}=(\overline{XX^T})_{\mathscr A}\beta+(\overline{XZ^T\theta})_{\mathscr A} \ \mbox{ and } \ (\overline{ZY})_{\mathscr A}=(\overline{ZX^T})_{\mathscr A}\beta+(\overline{ZZ^T\theta})_{\mathscr A}.\end{eqnarray} In the above two equations, however, the number of unknown parameters $\beta$ and $\theta_i$ with $i\in \mathscr A$ is larger than sample size $|\mathscr A|$. We then need an approach to reducing the number of unknowns.
It can be seen that
\begin{eqnarray*}E\left[(\overline{XZ^T\theta})_{\mathscr A}\right]=E[XZ^T]\overline{\theta}_{\mathscr A} \ \mbox{ and } \
Cov\left[(\overline{XZ^T\theta})_{\mathscr A}\right]=\frac{1}{|\mathscr A|}(\overline {\theta^TW_1\theta})_{\mathscr A},\end{eqnarray*} where $W_1=E[(ZX^T-E[ZX^T])(XZ^T-E[XZ^T])]$. As a result,
\begin{eqnarray*}(\overline{XZ^T\theta})_{\mathscr A}=E[XZ^T]\overline{\theta}_{\mathscr A} +O_p\left(1/\sqrt{|\mathscr A|}\right).\end{eqnarray*}  Similarly, we have
\begin{eqnarray*}
(\overline{ZZ^T\theta})_{\mathscr A}=E[ZZ^T]\overline{\theta}_{\mathscr A} +O_p\left(1/\sqrt{|\mathscr A|}\right).
\end{eqnarray*} Thus, the estimating equations in (\ref{(2.1)}) can be approximately expressed as
\begin{eqnarray}\label{(2.2)}(\overline{XY})_{\mathscr A}=(\overline{XX^T})_{\mathscr A}\beta+(\overline{XZ^T})_{\mathscr A}\overline{\theta}_{\mathscr A} \ \mbox{ and } \ (\overline{ZY})_{\mathscr A}=(\overline{ZX^T})_{\mathscr A}\beta+(\overline{ZZ^T})_{\mathscr A}\overline{\theta}_{\mathscr A}.\end{eqnarray} By solving the equations above, when $\mathscr A$ is chosen as $\mathscr G$, the estimator of $\beta$ is obtained as
\begin{eqnarray}\label{(2.3)}\widehat\beta=\left(\overline{X X^T}-\overline{XZ^T}(\overline{ZZ^T})^{-1}\overline{Z X^T}\right)^{-1}\left(\overline{XY}-\overline{XZ^T}(\overline{ZZ^T})^{-1}\overline{ZY}\right),\end{eqnarray}
and generally, the estimator of a subgroup-average $\overline\theta_{\mathscr A}$ is attained as
\begin{eqnarray}\label{(2.4)}\widehat{\overline\theta}_{\mathscr A}=\left((\overline{ZZ^T})_{\mathscr A}\right)^{-1}\left((\overline{ZY})_{\mathscr A}-(\overline{Z X^T})_{\mathscr A}\widehat\beta\right).\end{eqnarray}

{\bf Remark 2.1.} {\it (1) Theorem 3.1 given in the next section ensures that the estimators in (\ref{(2.3)}) and (\ref{(2.4)}) are consistent estimators of $\beta$ and ${\overline\theta}_{\mathscr A}$, respectively. Particularly, if the parameters $\theta_i$ with $i\in\mathscr G_j\subset \mathscr G$ are identical to $\theta_j^0$, then $\widehat{\overline\theta}_{\mathscr G_j}$ is the consistent estimator of $\theta^0_j$.
(2) It can be seen that the constructions of above estimators are free of the sparsity condition. Also Theorem 3.1 shows that the consistency of the estimators is free of the sparsity condition.}

\subsection{Classification}

Although the consistent estimators proposed above do not need the sparsity condition, for classification consistency, we require the following assumption:
\begin{itemize}\item[{\it C1}.] The index set $\mathscr G=\{1,\cdots,n\}$ can be separated into subgroups $\mathscr G_j,j=1,\cdots,k,$ such that
$$\theta_{i}=\theta_j^0 \mbox{ for } i\in\mathscr G_j,\ \bigcup_{j=1}^k\mathscr G_j=\mathscr G \ \mbox{ and } \ \mathscr G_{j_1}\bigcap \mathscr G_{j_2}=\emptyset \ \mbox{ for }j_1\neq j_2,$$ where the subgroups $\mathscr G_j$ and the number $k$ of the subgroups are unknown in advance. Moreover, it is supposed that $m_j\rightarrow\infty$ as $n\rightarrow\infty$, where $m_j=|\mathscr G_j|$ is the size of $\mathscr G_j$, i.e., the number of elements of $\mathscr G_j$.  \end{itemize} This condition is commonly used in the literature on clustering and classification. Note that here the number $k$ of the subgroups may tend to infinity as $n$ goes to infinity. The value of $k$ characters the sparsity level of the index set $\mathscr G$. More precisely, the larger value the number $k$ has, the more sparse the index set $\mathscr G$ is. More detailed condition will be given in the condition {\it C6} in the next section.

\subsubsection{Convex clustering}

If the homogenous parameter $\beta$ is given, it can bee seen from the model (\ref{(1.1)}) that the LS-based estimating equation for each heterogenous parameter $\theta_i$ is
\begin{eqnarray*}Z_i(Y_i-X_i^T\beta)=Z_iZ_i^T\theta_i.\end{eqnarray*} Note that $E[Z_i(Y_i-X_i^T\beta)]=\alpha_i$ with $\alpha_i=E[ZZ^T]\theta_i$, and identifying $\theta_i$ is equivalent to identifying $\alpha_i$. We then estimate the common values $\alpha_j^0=E[ZZ^T]\theta_j^{0}$ (i.e., identify the subgroups $\mathscr G_j$) by minimizing the following clustering criterion:
\begin{eqnarray}\label{(2.5)}\frac{1}{2}\sum_{i=1}^{n}
\left\|Z_i(Y_i-X_i^T\widehat\beta)-\alpha_i\right\|_2^2+\sum_{i<j}p_\gamma(\|\alpha_i-\alpha_j\|_1,\lambda),\end{eqnarray} where $\|\cdot\|_{p}$ is $L_p$-norm and $p_\gamma(t,\lambda)$ is a given penalty function. Here the penalty is used to encourage the sparsity in the differences between $\alpha_i$ and $\alpha_j$, i.e. flatness of the coefficient profiles $\alpha_i$ as a function of $i$.
In this paper, we only consider the MCP penalty (Zhang (2010)) defined as
$$p_\gamma(t,\lambda)=\lambda\int_0^t(1-x/(\gamma\lambda))_+dx, $$ where $\lambda$ is the Lagrange multiplier, the parameter $\gamma$ controls the concavity of the penalty function and is chosen as $\gamma>1$.

Denote by $\widetilde{\alpha}_1,\cdots,\widetilde{\alpha}_n$ the estimators of ${\alpha}_1,\cdots,{\alpha}_n$ defined by minimizing (\ref{(2.5)}) and call them the pairwise fusion estimators. Then, the pairwise fusion estimators of ${\theta}_1,\cdots,{\theta}_n$ can be expressed as
\begin{eqnarray}\label{(2.6)}\widetilde{\theta}_1=\overline{ZZ^T}^{-1}\widetilde{\alpha}_1,\cdots,
\widetilde{\theta}_n=\overline{ZZ^T}^{-1}\widetilde{\alpha}_n.\end{eqnarray}

{\bf Remark 2.2.} {\it It is worth pointing out that the regression coefficient $\beta$ is consistently estimated before clustering, and the covariate $Z_i$ is absorbed into $\alpha_i$ (i.e., $\theta_i$ is replaced by $\alpha_i$). Thus, the resulting criterion (\ref{(2.5)})
is a standard convex clustering (see, e.g., Chi and Lange; 2014), and the related algorithm is much easier than that designed for regression (see Ma and Huang; 2016b). Moreover, the algorithm only involves the $d_Z$-dimensional parameters, implying that it is stable; for the details see the next subsection.}

\subsubsection{Algorithm}

We use the alternating direction method of
multipliers (ADMM) to tackle the task of computation (Chi and Lange; 2014). To this end, we first recast the objective function (\ref{(2.5)}) as the constrained objective function:
\begin{eqnarray}\label{(2.7)}\frac{1}{2}\sum_{i=1}^{n}\left(Z_i(Y_i-X_i^T\widehat\beta)-\alpha_i\right)^2+
\sum_{i<j}p_\gamma(\|v_{ij}\|_1,\lambda) \ \mbox{ s.t. } \alpha_i-\alpha_j-v_{ij}=0. \end{eqnarray} The augmented Lagrangian for criterion (\ref{(2.7)}) is given by
\begin{eqnarray}\label{(2.8)}\nonumber &&\frac{1}{2}\sum_{i=1}^{n}\left\|Z_i(Y_i-X_i^T\widehat\beta)-\alpha_i\right\|_2^2+\sum_{i<j}
p_\gamma(\|v_{ij}\|_1,\lambda)\\&&+ \sum_{i<j}\kappa^T_{ij}(\alpha_i-\alpha_j-v_{ij})+\frac{\eta}{2}\sum_{i<j}\|\alpha_i-\alpha_j-v_{ij}\|_2^2, \end{eqnarray} where the dual variable vectors $\kappa_{ij}$ are Lagrange multipliers and $\eta$ is the penalty parameter. The objective function (\ref{(2.8)}) is convex with respect to each $v_{ij}$ when $\gamma>1/\eta$.

Write ${\bbf \alpha}=(\alpha_1,\cdots,\alpha_n)^T$, ${\bf v}=(v_{ij}:i<j)^T$ and ${\bbf\kappa}=(\kappa_{ij}:i<j)^T$. For given $\bf v$ and $\bbf\kappa$, an equivalent form of (\ref{(2.8)}) is
\begin{eqnarray*}L({\bbf\alpha}|{\bf v},{\bbf\kappa})=\frac{1}{2}\sum_{i=1}^{n}\left\|Z_i(Y_i-X_i^T\widehat\beta)-\alpha_i\right\|_2^2
+\frac{\eta}{2}\sum_{i<j}\|\alpha_i-\alpha_j-\widetilde v_{ij}\|_2^2+C,\end{eqnarray*} where $\widetilde v_{ij}=
v_{ij}-\eta^{-1}\kappa_{ij}$, and $C$ is independent of $\bbf\alpha$.
Take the derivative of $L({\bbf\alpha}|{\bf v},{\bbf\kappa})$ with respect to $\alpha_i$ and let the derivative to be zero. We get the analytical solution of $\alpha_i$ as
\begin{eqnarray}\label{(2.9)}\alpha_i=\frac{1}{1+n\eta}W_i+\frac{n\eta}{1+n\eta}
\frac{1}{n}\sum_{i=1}^{n}Z_i(Y_i-X_i^T\widehat\beta),\end{eqnarray}
where $$W_i=Z_i(Y_i-X_i^T\widehat\beta)+\sum_{j\neq i,1\leq j\leq n}(\kappa_{ij}+\eta^{-1} v_{ij})-\sum_{1\leq j<i}(\kappa_{ji}+\eta^{-1} v_{ji}).$$ Then, the algorithm can be separated into the following steps:

{\it Step 1.} Choose the
initial values $\bbf \kappa^{(0)}$ and ${\bf v^{(0)}}$ of $\bbf \kappa$ and ${\bf v}$, respectively.

{\it Step 2.} For $m=1,2\cdots,$ and $i=1,\cdots,n$, calculate \begin{eqnarray*}W_i&=&Z_i(Y_i-X_i^T\widehat\beta)+\sum_{j\neq i,1\leq j\leq n}(\kappa_{ij}^{(m-1)}+\eta^{-1} v_{ij}^{(m-1)})\\&&-\sum_{1\leq j<i}(\kappa_{ji}^{(m-1)}+\eta^{-1} v_{ji}^{(m-1)}).\end{eqnarray*}

{\it Step 3.}  For $i=1,\cdots,n$, update $\alpha_i$ as $${\alpha_i^{(m)}}=\frac{1}{1+n\eta}W_i+\frac{n\eta}{1+n\eta}\frac{1}{n}\sum_{i=1}^{n}Z_i(Y_i-X_i^T\widehat\beta).$$

{\it Step 4.}  For $j=1,\cdots,n$, update $v_{ij}$ and $\kappa_{ij}$ respectively as
\begin{eqnarray*}\left\{\begin{array}{cc}
v^{(m)}_{ij}=\left\{\begin{array}{cc}\frac{\varsigma^{(m)}_{ij}}{1-1/(\gamma\eta)}& \mbox{ if } \
\|\delta_{ij}\|_1\leq \gamma\lambda\\\delta_{ij} & \mbox{ if }\
\|\delta_{ij}\|_1> \gamma\lambda,\end{array}\right. \vspace{2ex}
\\ \kappa^{(m)}_{ij}=\kappa^{(m-1)}_{ij}
+\eta\left(v^{(m)}_{ij}-{\alpha_i^{(m)}}+{\alpha_j^{(m)}}\right),\end{array}\right.\end{eqnarray*} where $\delta_{ij}={\alpha_i^{(m)}}-{\alpha_j^{(m)}}+
\eta^{-1}\kappa_{ij}^{(m-1)}$ and
$$\varsigma^{(m)}_{ij}=\argmin_{v_{ij}}\frac12\left[\left\|v_{ij}-\left({\alpha_i^{(m)}}-{\alpha_j^{(m)}}-
\eta^{-1}\kappa_{ij}^{(m-1)}\right)\right\|_2^2
+\eta^{-1}\lambda\|v_{ij}\|_1\right].$$

{\it Step 5.} Terminate the algorithm if the stopping rule is met at step $m+1$. Then, ${\alpha_i^{(m+1)}}$ are the final choices. Otherwise, repeat the above steps.

{\it Step 6.} After identifying $\mathscr G_j$, the final estimator of $\theta_j^0$ is the following subgroup-average estimator:
$$\widehat{\overline\theta}_{\mathscr G_j}=\left((\overline{ZZ^T})_{\mathscr G_j}\right)^{-1}\left((\overline{ZY})_{\mathscr G_j}-(\overline{Z X^T})_{\mathscr G_j}\widehat\beta\right).$$

\

For the calculation steps, we have following explanations:

{\bf Remark 2.3.} {\it(1) The proposed estimators and algorithm depend on the tuning parameter $\lambda$. We can estimate it by the commonly used methods such as CV given in Fan and Li (2001). (2) Like the convergence of the ADMM, the convergence of the above algorithm
can be guaranteed for any $\eta>0$; for details see Chi and Lange (2014). (3) The algorithm above only involves the iterations of the $d_Z$-dimensional parameters. Thus, it is simple and stable.}

\setcounter{equation}{0}
\section{Theoretical properties}

We first establish the asymptotic normality for regression coefficient estimator (\ref{(2.3)}) and subgroup-average estimator (\ref{(2.4)}). Because of the heterogeneity of $\theta_i,i=1,\cdots,n$, we need the following controllability condition to get certain theoretical conclusions.

\begin{itemize}\item[{\it C2}.] As $n\rightarrow\infty$, $|\mathscr A|\rightarrow\infty$ and
\begin{eqnarray*}\frac{1}{n}\sum_{i=1}^n\left(\theta_i-
\overline\theta\right)W_1\left(\theta_i-
\overline\theta\right)^T\rightarrow \Psi \ \mbox{ and } \
\frac{1}{|\mathscr A|}\sum_{i\in\mathscr A}\left(\theta_i-\overline\theta_{\mathscr A}\right)W_2\left(\theta_i-\overline\theta_{\mathscr A}\right)^T\rightarrow \Lambda_{\mathscr A},\end{eqnarray*} where \begin{eqnarray*}W_1&=&E\{(Z X^T-E[Z X^T])(Z X^T-E[Z X^T])^T\}\\W_2&=&E\{(ZZ^T-E[ZZ^T]-E[Z X^T]\Omega^{-1}(XZ^T-E[XZ^T]))\\&&\times(ZZ^T-E[ZZ^T]-E[Z X^T]\Omega^{-1}(XZ^T-E[XZ^T]))^T\}\end{eqnarray*} with $\Omega=E[X X^T]-E[XZ^T](E[ZZ^T])^{-1}E[Z X^T]$, and $\Psi$ and $\Lambda_{\mathscr A}$ are some positive definite matrices.\end{itemize}

This condition means that the level of the heterogeneity of $\theta_i$ should be in a certain range. Intuitively, the condition is similar to the notion that a random variable has a finite variance. The condition is common, and under the sparsity condition {\it C1} with small $k$, for example, the condition is satisfied. Write \begin{eqnarray*}&&\Phi=Var\left[X-E[XZ^T](E[ZZ^T])^{-1}Z\right], \\&& \Upsilon=Var\left[Z-\Omega^{-1}\left(X-E[XZ^T](E[ZZ^T])^{-1}Z\right)\right],\end{eqnarray*} and
denote by $\beta^0$ and $\overline\theta^0_{\mathscr A}$ the true values of $\beta$ and $\overline\theta_{\mathscr A}$ respectively. The following theorem states the asymptotic normality.

\noindent{\bf Theorem 3.1.} {\it Under condition C2, if $E[XX^T]$ and $E[ZZ^T]$ exist, then the regression coefficient estimator (\ref{(2.3)}) and subgroup-average estimator (\ref{(2.4)}) respectively have the following asymptotic properties:
\begin{eqnarray*}&&\sqrt n(\widehat\beta-\beta^0)\stackrel{D}\longrightarrow N\left(0,\Omega^{-1}(\Psi+\sigma^2\Phi)\Omega^{-1}\right),\\&&\sqrt{|\mathscr A|}\Big(\widehat{\overline\theta}_{\mathscr A}-\overline\theta^0_{\mathscr A}\Big)\longrightarrow N\left(0,(E[ZZ^T])^{-1}(\Lambda_{\mathscr A}+\sigma^2\Upsilon)(E[ZZ^T])^{-1}\right).\end{eqnarray*}
}

For the theorem, we have the following explanations.

\noindent{\bf Remark 3.1.} {\it 1) The consistency of the two estimators is free of the sparsity level $k$ given in {\it C1}. However, the existing methods such as penalty-based methods cannot achieve the estimation consistency if without the sparsity condition (see, e.g., Lin et al.; 2016, Ma and Huang; 2016, and Guo et al.; 2016).

2) From the theorem we see that when $\Psi$ and $\Lambda_{\mathscr A}$ are large, the two estimators have a large estimation variance, implying that a strong heterogeneity of $\theta_i$ can reduce the estimation efficiency.

}

\

In the following, we focus on the asymptotic property of the pairwise fusion estimators $\widetilde{\theta}_1,\cdots,\widetilde{\theta}_n$ defined by (\ref{(2.6)}). This is the most key  for classification. To this end, we first introduce the following notations. Suppose without loss of generality that the subgroups $\mathscr G_j,j=1,\cdots,k$, are orderly separations of $\mathscr G$ such that
$\mathscr G=(\mathscr G_1,\cdots,\mathscr G_k)$ with
$\mathscr G_j=(j_1,\cdots,j_{m_j})$. Denote ${\bbf\alpha}^0=\left({\alpha_1^0}^T,\cdots,{\alpha_k^0}^T\right)^T$ with $\alpha_j^0=E[ZZ^T]\theta_j^0$,
${\bf Y}=(Y_{1},\cdots,Y_{n})^T$, ${\bf X}=({\bf x}_1,\cdots,{\bf x}_{d_X})=(X_{1},\cdots,X_{n})^T$, ${\bf Z}=({\bf z}_1,\cdots,{\bf z}_{d_Z})=(Z_{1},\cdots,Z_{n})^T$ and  ${\mathscr I}=\mbox{diag}({\bf 1}_1,\cdots,{\bf 1}_k)$, where ${\bf 1}_j=(1,\cdots,1)^T$, an $m_j$-dimensional vector with all the elements equal to 1.
For any vector $\zeta=(\zeta_1,\cdots,\zeta_m)^T$, denote $\|\zeta\|_\infty=\max_{1\leq l\leq m}|\zeta_l|$, and for any matrix $A=(a_{ij})_{i=1,j=1}^{s,t}$ denote $\|A\|_\infty=\max_{1\leq i\leq s}\sum_{j=1}^t|a_{ij}|$. Write $|\mathscr G|_{\min}=\min\limits_{j=1,\cdots,k}\{m_j\}$ and $|q|_{\max}=\max\{(\sum_{j=1}^nq_{ij}^2)^{1/2}:i=1,\cdots,kd_Z\}$, where $q_{ij}$ is the $(i,j)$-th element of the $(kd_Z)\times n$ matrix $\left(E[XZ^T],\cdots,E[XZ^T]\right)^TE[{\bf Q}]$ with $${\bf Q}=\left({\bf X}^T{\bf X}-{\bf X}^T {\bf Z}({\bf Z}^T{\bf Z})^{-1}{\bf Z}^T {\bf X}\right)^{-1}\left({\bf X}^T-{\bf X}^T{\bf Z}({\bf Z}^T{\bf Z})^{-1}{\bf Z}^T\right).$$

If the group structure in {\it C1} was known beforehand, the oracle estimators of $\alpha_j,j=1,\cdots,k$, are defined as
\begin{eqnarray}\label{(3.1)}\widehat\alpha_i^o=\widehat\alpha_j^0  \ \mbox{ for } i\in\mathscr G_j,\end{eqnarray} where
\begin{eqnarray}\label{(3.2)}\left(\widehat\alpha_1^0,\cdots,\widehat\alpha_k^0\right)=
\argmin_{\alpha_j^0\in\Theta,j=1,\cdots,k}L({\bbf \alpha}^0)\end{eqnarray} with
$L({\bbf \alpha}^0)=\frac{1}{2}\sum_{j=1}^k\sum_{i\in\mathscr G_j}\left\|Z_i(Y_i-X_i^T\widehat\beta)-\alpha_j^0\right\|_2^2$ and  $\Theta$ being the parameter space of $\alpha_i$.
The estimators in (\ref{(3.2)}) can be further rewritten as
\begin{eqnarray*}\left(\widehat\alpha_1^0,\cdots,\widehat\alpha_k^0\right)=
\argmin_{{\bbf\alpha}^0\in\Theta}\frac{1}{2}\Big\|{\bf Z}\circ({\bf Y}-
{\bf X}\widehat\beta)-{\mathscr I}{\bbf\alpha}^0\Big\|_2^2,\end{eqnarray*} where $``\circ"$ denotes the Hadamard product. Denote \begin{eqnarray*}&&
V_{ij}=\sigma^2 (1-b_{i})
(1-b_{j})Var(X-E[XZ^T](E[ZZ^T])^{-1}Z),\\&& V_j=\sigma^2 Var\{Z+b_{j}E[XZ^T]\Omega^{-1}(X-E[XZ^T](E[ZZ^T])^{-1}Z)\},\\&&\Gamma_j=V_j+
b_{j}E[ZX^T]\Omega^{-1}(\Psi+\sigma^2(1-b_{j})\Phi)\Omega^{-1}E[XZ^T].\end{eqnarray*} For the oracle estimators, we have the following asymptotic normality.

\noindent{\bf Lemma 3.2.} {\it If condition {\it C1} and the conditions of Theorem 3.1 hold, then, for any vector ${\bf a}_n\in R^{kp}$, the oracle estimators have the following asymptotical normality:
\begin{eqnarray*}&&({\bf a}_n^T(\sigma^2\mathscr D+\mathscr V(\theta)){\bf a}_n)^{-1/2}{\bf a}_n^T\left(\sqrt m_1(\widehat\alpha_i^o-\alpha_i^o)^T_{i\in\mathscr G_1},\cdots,\sqrt m_k(\widehat\alpha_i^o-\alpha_i^o)^T_{i\in\mathscr G_k}\right)^T\hspace{1ex}\\&&\stackrel{D}\longrightarrow
N\left(0,1\right),\end{eqnarray*} where the $(kd_Z)\times (kd_Z)$ matrix $\mathscr D=(\Delta_{ij})$ with $\Delta_{jj}=\Gamma_j$ and $\Delta_{ij}=\Omega^{-1}(\Psi+
V_{ij})\Omega^{-1}$ for $i\neq j$ and $i,j=1,\cdots,k$, and the $(kd_Z)\times (kd_Z)$ matrix $\mathscr V(\theta)=diag(Cov(\theta^TZZ^T),\cdots,Cov(\theta^TZZ^T))$.
}

From the lemma, we have the following findings.

\noindent{\bf Remark 3.2.} {\it Although the oracle estimator $\widehat\alpha_i^o$ is consistent, unlike the estimator for the homogeneous parameter $\beta$, which has the standard convergence rate of order $\sqrt n$, the oracle estimator $\widehat\alpha_i^o$ has a slower convergence rate of order $\sqrt m_j$ for some $j$. It is because, as shown by the proof of the theorem, actually the oracle estimator $\widehat\alpha_i^o$ ($i\in\mathscr G_j$) is constructed mainly by the data with indices in $\mathscr G_j$.
}

In order to establish the related asymptotic theory, we need the following conditions:
\begin{itemize} \item[{\it C3.}] $\|{\bf x}_i\|_2=\sqrt{n}$ and $\|{\bf z}_j\|_2= \sqrt{n}$ for all $i=1,\cdots,d_X,j=1,\cdots,d_Z$, and $\sqrt{n/|q|_{\max}}\geq c_1$ when $n$ is large enough, where the constant $c_1>0$.
\item[{\it C4.}] The error vector ${\bbf\epsilon}=(\epsilon_1,\cdots,\epsilon_n)^T$ of model (\ref{(1.1)}) has sub-Gaussian tails such that $P(|{\bf a}^T{\bbf\epsilon}|>\|{\bf a}\|_2x)\leq 2\exp(-c_2x^2)$ for any vector ${\bf a}\in R^n$ and $x>0$, where the constant $0<c_2<\infty$.
    \item[{\it C5}.] $E[\|Z\|_2^{4+2\delta}]<\infty$ for some constant $0<\delta<1$.
    \item[{\it C6}.] The sizes $m_j$ of $\mathscr G_j,j=1,\cdots,k$, satisfy $m_j=O(n^{1-\varepsilon_j})$ and $\sqrt{m_j/n}\rightarrow b_{j}$ as $n\rightarrow\infty$, where the constants $\varepsilon_1,\cdots,\varepsilon_k$ satisfy $0\leq\varepsilon_j<\min\{c_1,c_1c_2,1/2\}$ and $b_{j}\geq 0$.
\end{itemize}
The condition {\it C3} is mild clearly. The condition {\it C4} is commonly used in high-dimensional settings.
We need the condition {\it C5} together with the first assumption in the condition {\it C6} to guarantee the convergence rate in Central Limit Theorem (see, e.g., Osipov and Petrov 1967). The first assumption in the condition {\it C6} implies the sparsity of the index set $\mathscr G$.
The second assumption in the condition {\it C6} is required only for a certain covariance structure in the asymptotic normality given below.

Denote $\phi_n=|\mathscr G|_{\min}^{-1}\sqrt{n\log n}$ and \begin{eqnarray*}\delta_n&=&\max\Big\{2d_Xd_Z[(\sqrt{\log n})^{-1}n^{-1/2}+\varphi(\sqrt{|\mathscr G|_{\min}}(1+\log n) )|\mathscr G|_{\min}^{-\delta /2}(\sqrt{\log n})^{-2-\delta}],\\&&
\hspace{9.5cm} 2kd_Zn^{-c_2},2kd_Zn^{-c_1c_2}\Big\},\end{eqnarray*} where $c_1$ and $c_2$ are defined in {\it C1} and {\it C2} respectively, and $\varphi(u)$ is a certain function, defined in the region $u>0$, bounded and non-increasing with $\lim\limits_{u\rightarrow\infty}\varphi(u)=0$. We have the following lemma.

\

\noindent{\bf Lemma 3.3.} {\it Under the conditions C1-C6,
we have that with probability at least $1-\delta_n$,
$$\left\|((\widehat\alpha_1^0-\alpha_1^o)^T,\cdots,
(\widehat\alpha_k^0-\alpha_k^o)^T)^T\right\|_\infty\leq
\phi_n,$$ where $\alpha_j^o$ is the true value of $\alpha_j^0$. Consequently, the oracle estimators defined in (\ref{(3.1)}) satisfy
$$\left\|((\widehat\alpha_1^o-\alpha_1^o)^T,\cdots,
(\widehat\alpha_n^o-\alpha_n^o)^T)^T\right\|_\infty\leq
\phi_n.$$
}

From the condition {\it C6}, we see that $\phi_n\rightarrow 0$ and $\delta_n\rightarrow 0$ as $n\rightarrow\infty$, implying the oracle estimators are strongly consistent.
For further conclusion, we write
$$b_n=\min_{i\in\mathscr G_s,j\in\mathscr G_t,s\neq t}\|\alpha_i^0-\alpha_j^0\|.
$$ We have the following lemma.

\noindent{\bf Lemma 3.4.} {\it Under the conditions Lemma 3.2, if $b_n>\gamma\lambda$ and $\lambda\gg\phi_n$, then the pairwise fusion estimators $\widetilde{\alpha}_1,\cdots,\widetilde{\alpha}_n$ defined by minimizing (\ref{(2.5)}) satisfy
$$P\left(\widetilde{\alpha}_1=\widehat\alpha^{o}_1,\cdots,\widetilde{\alpha}_n=\widehat\alpha^{o}_n\right)\rightarrow 1  \mbox{ as } n\rightarrow\infty.$$
}

By combining the above lemmas, we get the following theorem.

\noindent{\bf Theorem 3.5.} {\it If the conditions of Lemma 3.4 hold, then, for any vector ${\bf a}_n\in R^{kp}$, the pairwise fusion estimators $\widetilde{\alpha}_1,\cdots,\widetilde{\alpha}_n$ have the following asymptotical normality:
\begin{eqnarray*}&&({\bf a}_n^T(\sigma^2\mathscr D+\mathscr V(\theta)){\bf a}_n)^{-1/2}{\bf a}_n^T\left(\sqrt m_1(\widetilde\alpha_i-\alpha_i^{o})^T_{i\in\mathscr G_1},\cdots,\sqrt m_k(\widetilde\alpha_i-\alpha_i^{o})^T_{i\in\mathscr G_k}\right)^T\hspace{1ex}\\&&\stackrel{D}\longrightarrow
N(0,1) \ \mbox{ as }n\rightarrow\infty.\end{eqnarray*}}

Finally, by the relation between $\widetilde{\theta}_1,\cdots,\widetilde{\theta}_n$ and $\widetilde{\alpha}_1,\cdots,\widetilde{\alpha}_n$, we attain the following corollary.

\noindent{\bf Corollary 3.6.} {\it If the conditions of Lemma 3.4 hold, then, for any vector ${\bf a}_n\in R^{kp}$, the pairwise fusion estimators ${\theta}_1,\cdots,{\theta}_n$ have the following asymptotical normality:
\begin{eqnarray*}&&({\bf a}_n^T(\Lambda(\sigma^2\mathscr D+\mathscr V(\theta))\Lambda){\bf a}_n)^{-1/2}{\bf a}_n^T\left(\sqrt m_1(\widetilde\theta_i-\theta_i^{o})^T_{i\in\mathscr G_1},\cdots,\sqrt m_k(\widetilde\theta_i-\theta_i^{o})^T_{i\in\mathscr G_k}\right)^T\hspace{1ex}\\&&\stackrel{D}\longrightarrow
N\left(0,1\right) \  \mbox{ as }n\rightarrow\infty,\end{eqnarray*}  where the $(kd_Z)\times (kd_Z)$ matrix $\Lambda=\mbox{diag}(E^{-1}[ZZ^T],\cdots,E^{-1}[ZZ^T])$.
}

\

Similar to Remark 3.2, we have the following remark.

\noindent{\bf Remark 3.3.} {\it The pairwise fusion estimator $\widetilde\theta_i$ for the heterogenous parameters $\theta_i$ ($i\in\mathscr G_j$) has a slower convergence rate of order $\sqrt m_j$. Actually the estimator depends mainly on data with indices in $\mathscr G_j$.  }

\

\setcounter{equation}{0}
\section{Numerical analyses}

\subsection{Simulation studies}
Now we conduct some simulation studies to examine the finite sample behavior of our method, and compare ours with the methods of Ma and Huang (2016b) and the ordinary least squares (OLS). For a comprehensive comparison, the empirical mean, median and standard deviation (std) are employed to valuate the behaviors of the estimators of $\theta_i$, $k$ and $\beta$. Moreover, the percentage (per) of accurately estimating $k$ is used to further measure the performances of the estimators of $k$. The simulation results for these empirical criteria are obtained via repeating the experiments 200 times. For the heterogenous models under study, the behaviors of the estimation and classification depend on the variability of the random variable $Z$. We then consider the simulations in the cases with different values of coefficient of variation, $\sigma/\mu$, where $\mu$ and $\sigma$ are the mean and standard deviation of a component of $Z$. In the simulation procedure, the tuning parameter is chosen by the CV criterion.

\textit{Example 1 (Single treatment effect)}. In this experiment, the data are generated from the heterogeneous model with univariate $Z$ as
\begin{equation*}
  Y_i = X_i^T \beta + Z_i\theta_i + \epsilon_i, i = 1,\cdots, n,
\end{equation*}
where random variables $X_i=(X_{i1}, X_{i2}, X_{i3})^T$ come from a 3-dimensional normal distribution with mean 0, variance 1 and correlation coefficient $\rho=0.3$, variables $Z_i$ follow the normal distribution with mean $\mu$ and variance $\sigma^2$, and the error terms $\epsilon_i\sim N(0,0.5^2)$. With different choices of pair $(\mu,\sigma)$, we may get the different values of coefficient of variation $\sigma/\mu$. In the procedure of simulation, the homogenous coefficients are set as
$\beta = (2,2,2)^T$, the heterogenous coefficients $\theta_i$ are randomly divided into two subgroups with equal probabilities, i.e., $P(j\in \mathcal{G}_1) = P(j\in \mathcal{G}_2) = 0.5$, and $\theta_j = \theta^0_1$ for $j\in \mathcal{G}_1$ and $\theta_j = \theta^0_2$ for $j\in \mathcal{G}_2$.
For different choices of $(\mu, \sigma)$, $\theta_j^0$ and sample size, the simulation results are reported in Table \ref{table1} and Table \ref{table2}.
From the two tables, we have the following conclusions.

(1) Generally speaking, with the decrease of coefficient of variation $\sigma/\mu$, our method has a great improvement. Contrarily, Ma's method performs badly for the case with small value of $\sigma/\mu$.

(2) For the case with large value of $\sigma/\mu$, the difference between ours and Ma's is not significant.

(3) It is clear that the estimation and classification of our method are more robust than those of Ma's method under the criterion of standard deviation (std).

\begin{table}[htbp]
\centering
\caption{Empirical mean, median and standard deviation of the estimator of $k$, and the percentage of estimated $k$ equaling to the true member of subgroups in Example 1 \label{table1}}
\small
\begin{tabular}{cll|cccc|cccc}
  \hline
  &&   &\multicolumn{4}{c|}{$\theta^0_1=1,\theta^0_2=-1$} &\multicolumn{4}{c}{$\theta^0_1=2,\theta^0_2=-2$} \\ \hline
  $n$&$(\mu,~\sigma)$&Methods&mean&median&std&per&mean&median&std&per\\ \hline
  100&(3.0,1.0)&Ours&1.920 &2.000 &0.484 &0.760 &2.000 &2.000 &0.425 &0.820 \\
     &         &Ma's&1.995 &2.000 &0.506 &0.745 &2.050 &2.000 &0.498 &0.770 \\
     &(2.0,0.5)&Ours&2.010 &2.000 &0.100 &0.860 &1.945 &2.000 &0.228 &0.945 \\
     &         &Ma's&1.700 &2.000 &0.594 &0.560 &1.930 &2.000 &0.612 &0.670 \\
     &(1.0,0.2)&Ours&2.080 &2.000 &0.338 &0.880 &2.080 &2.000 &0.273 &0.920 \\
     &         &Ma's&2.290 &2.000 &1.112 &0.460 &1.890 &2.000 &0.567 &0.670 \\ \hline
  200&(3.0,1.0)&Ours&1.780 &2.000 &0.506 &0.820 &2.180 &2.000 &0.388 &0.820 \\
     &         &Ma's&1.860 &2.000 &0.452 &0.780 &2.040 &2.000 &0.283 &0.920  \\
     &(2.0,0.5)&Ours&1.840 &2.000 &0.468 &0.880 &1.960 &2.000 &0.284 &0.980 \\
     &         &Ma's&2.440 &2.000 &1.052 &0.520 &1.840 &2.000 &0.370 &0.840  \\
     &(1.0,0.2)&Ours&2.080 &2.000 &0.488 &0.820 &1.960 &2.000 &0.282 &0.980  \\
     &         &Ma's&3.240 &3.000 &1.064 &0.480 &3.140 &3.000 &1.224 &0.440 \\\hline
\end{tabular}
\end{table}

\begin{table}[htbp]\small
\centering
\caption{Empirical mean, median and standard deviation of estimators $\theta^0_1$ and $\theta^0_2$ in Example 1 \label{table2}}
\begin{tabular}{ccl|ccc|ccc}
  \hline
  $n$    &$(\mu,~\sigma)$&Methods&mean&median&std&mean&median&std \\ \hline
  &&&\multicolumn{3}{c|}{$\theta^0_1=1$}&\multicolumn{3}{c}{$\theta^0_2=-1$} \\ \hline
  $100$  &(3.0,1.0)&Ours & 0.763 & 0.897 & 0.935 & -0.780 & -0.959 & 0.881 \\
         &         &Ma's & 0.711 & 0.812 & 0.379 & -0.845 & -1.026 & 0.871 \\
         &(2.0,0.5)&Ours & 0.996 & 1.002 & 0.039 & -1.025 & -1.027 & 0.040 \\
         &         &Ma's & 0.654 & 0.735 & 0.728 & -0.717 & -1.256 & 1.252 \\
         &(1.0,0.2)&Ours & 0.691 & 0.826 & 0.512 & -0.886 & -1.070 & 0.670 \\
         &         &Ma's & 0.680 & 0.709 & 1.146 & -0.686 & -1.281 & 1.890 \\ \hline
  $200$  &(3.0,1.0)&Ours & 0.987 & 0.987 & 0.032 & -0.989 & -0.988 & 0.034 \\
         &         &Ma's & 0.630 & 0.968 & 1.273 & -0.821 & -0.986 & 0.384 \\
         &(2.0,0.5)&Ours & 0.938 & 0.988 & 0.304 & -0.946 & -0.997 & 0.288 \\
         &         &Ma's & 0.882 & 0.963 & 0.395 & -0.907 & -0.979 & 0.387 \\
         &(1.0,0.2)&Ours & 0.564 & 0.872 & 0.816 & -0.726 & -1.051 & 0.680 \\
         &         &Ma's & 0.986 & 0.971 & 0.901 & -0.695 & -0.964 & 0.505 \\ \hline
  &&&\multicolumn{3}{c|}{$\theta^0_1=2$}&\multicolumn{3}{c}{$\theta^0_2=-2$} \\ \hline
  $100$  &(3.0,1.0)&Ours &1.667 &2.002 &1.034 &-1.354 &-1.493 &0.721 \\
         &         &Ma's &1.893 &1.947 &0.699 &-1.922 &-1.943 &0.222 \\
         &(2.0,0.5)&Ours &1.668 &1.941 &0.959 &-1.737 &-1.992 &0.978 \\
         &         &Ma's &1.657 &1.757 &0.814 &-1.607 &-1.964 &1.342 \\
         &(1.0,0.2)&Ours &1.916 &2.047 &0.681 &-1.852 &-1.964 &0.405 \\
         &         &Ma's &1.537 &2.511 &2.404 &-1.214 &-1.230 &1.382 \\ \hline
  $200$  &(3.0,1.0)&Ours &1.786 &1.991 &0.873 &-1.807 &-2.006 &0.878 \\
         &         &Ma's &1.945 &1.967 &1.105 &-1.701 &-1.952 &0.876 \\
         &(2.0,0.5)&Ours &1.994 &1.999 &0.027 &-1.993 &-1.994 &0.024 \\
         &         &Ma's &1.842 &1.944 &0.642 &-1.893 &-1.950 &0.358 \\
         &(1.0,0.2)&Ours &2.012 &2.021 &0.049 &-1.995 &-1.986 &0.062 \\
         &         &Ma's &1.965 &1.964 &0.037 &-1.880 &-1.888 &0.045 \\
  \hline
\end{tabular}
\end{table}

\clearpage

\textit{Example 2 (Two treatment effects)}. In this experiment, the data are generated from the heterogeneous model with 2-dimensional heterogenous coefficients as
\begin{equation*}
  Y_i = X_i^T \beta + Z_i^T\theta_i + \epsilon_i, i = 1,\cdots, n.
\end{equation*}
The other experiment condition are the same as those as in Experiment 1 except that $Z_i$ follow a 2-dimensional normal distribution as
\begin{equation*}
  Z_i \sim N \left\{\mu\left(\begin{array}{c} 1 \\ 1 \\ \end{array} \right),
  \sigma \left( \begin{array}{cc} 1 & 0.3 \\ 0.3 & 1 \\ \end{array} \right)
  \right\}.
\end{equation*}
Also for the heterogenous coefficients $\theta_j$, we randomly divide them into two subgroups with equal probabilities, i.e., $P(j\in \mathcal{G}_1) = P(j\in \mathcal{G}_2) = 0.5$, and $\theta_j = (\theta^0_{11}, \theta^0_{12})^T$ for $j\in \mathcal{G}_1$ and $\theta_j = (\theta^0_{21}, \theta^0_{22})^T$ for $j\in \mathcal{G}_2$.
We consider two different cases as follows:

\textit{case 1:} $(\theta^0_{11}, \theta^0_{12})^T = (1, 1)^T$ and
$(\theta^0_{21}, \theta^0_{22})^T = (-1, -1)^T$;

\textit{case 2:} $(\theta^0_{11}, \theta^0_{12})^T = (2, 2)^T$ and
$(\theta^0_{21}, \theta^0_{22})^T = (-2, -2)^T$.

\noindent The simulations results are presented in Table \ref{table3}, Table \ref{table4} and Table \ref{table5}. For each term in Table \ref{table4} and Table \ref{table5}, the above numerals are the simulation results of $(\theta^0_{11}, \theta^0_{12})$ and the bottom numerals are the simulation results of $(\theta^0_{21}, \theta^0_{22})$. Besides, we only display the standard deviation (std) of each component of estimators of $(\theta^0_{11}, \theta^0_{12})$ and $(\theta^0_{21}, \theta^0_{22})$, but the numeral results for the covariance between them are neglected. We have the following findings.

(1) By comparing the numerical behaviors of our method with those of Ma's displayed in Table \ref{table3}, Table \ref{table4} and Table \ref{table5}, we get the same comparative conclusions for both methods as in Experiment 1.

(2) On the other hand, by comparing the numerical results in Table \ref{table3}, Table \ref{table4} and Table \ref{table5} with those in Table \ref{table1} and Table \ref{table2}, we can see that the behaviors of our method are robust to the change of the dimension of $Z$ in the selected range.

(3) Moreover, by comparing the simulation results in case 1 with those in case 2, we can see that our method is robust to choices of the value of each component of $\theta_j^0$ in the selected range.

\begin{table}[htbp]
\small
\centering
\caption{Empirical mean, median and standard deviation of estimators of $k$ and the percentage of $k$ equaling to the true member of subgroups in Example 2. \label{table3}}
\begin{tabular}{cll|cccc|cccc}
  \hline
  &&   &\multicolumn{4}{c|}{\textit{case 1}} &\multicolumn{4}{c}{\textit{case 2}} \\ \hline
  $n$&$(\mu,~\sigma)$&Methods&mean&median&std&per&mean&median&std&per\\ \hline
  100&(3.0,1.0)&Ours& 1.980 & 2.000 & 0.246 & 0.940 & 2.100 & 2.000 & 0.303 & 0.900 \\
     &         &Ma's& 1.880 & 2.000 & 0.689 & 0.520 & 2.360 & 2.000 & 0.525 & 0.600 \\
     &(2.0,0.5)&Ours& 2.020 & 2.000 & 0.141 & 0.980 & 2.180 & 2.000 & 0.388 & 0.820 \\
     &         &Ma's& 2.160 & 2.000 & 0.833 & 0.480 & 2.600 & 3.000 & 0.857 & 0.480 \\
     &(1.0,0.2)&Ours& 2.020 & 2.000 & 0.141 & 0.980 & 2.120 & 2.000 & 0.328 & 0.880 \\
     &         &Ma's& 2.500 & 2.000 & 0.953 & 0.380 & 2.720 & 3.000 & 0.809 & 0.400 \\ \hline
  200&(3.0,1.0)&Ours& 2.000 & 2.000 & 1.030 & 0.500 & 2.040 & 2.000 & 0.988 & 0.720 \\
     &         &Ma's& 2.120 & 2.000 & 0.627 & 0.600 & 2.240 & 2.000 & 0.771 & 0.680 \\
     &(2.0,0.5)&Ours& 2.060 & 2.000 & 0.239 & 0.940 & 2.020 & 2.000 & 0.141 & 0.980 \\
     &         &Ma's& 2.160 & 2.000 & 0.738 & 0.640 & 2.260 & 2.000 & 0.560 & 0.680 \\
     &(1.0,0.2)&Ours& 2.020 & 2.000 & 0.141 & 0.980 & 2.340 & 2.000 & 0.478 & 0.660 \\
     &         &Ma's& 1.940 & 2.000 & 0.793 & 0.500 & 2.860 & 3.000 & 0.989 & 0.460 \\ \hline
\end{tabular}
\end{table}

\begin{table}[htbp]\small
\centering
\caption{Empirical mean, median and standard deviation of estimators of $(\theta^0_{11},\theta^0_{12})$ and $(\theta^0_{21},\theta^0_{22})$ for the \textit{case 1} in Example 2. \label{table4}}
\begin{tabular}{llllll}
  \hline
  $n$&$(\mu,\sigma)$&Methods&mean&median&std \\ \hline
  100&(3.0,1.0)&Ours&(0.682,0.678)  &(0.990,0.946)  &(0.689,0.704) \\
     &         &    &(-0.719,-0.709)&(-0.994,-0.958)&(0.745,0.659) \\ \cline{3-6}
     &         &Ma's&(1.444,0.639)  &(0.621,1.027)  &(1.182,1.449) \\
     &         &    &(-0.508,-1.135)&(-0.666,-1.165)&(1.827,1.729) \\ \cline{2-6}
     &(2.0,0.5)&Ours&(0.767,0.736)  &(0.955,0.974)  &(0.622,0.722) \\
     &         &    &(-0.754,-0.780)&(-0.996,-0.990)&(0.677,0.661) \\ \cline{3-6}
     &         &Ma's&(0.537,0.576)  &(0.673,0.836)  &(0.747,0.587) \\
     &         &    &(-0.698,-0.481)&(-0.856,-0.903)&(1.760,1.379) \\ \cline{2-6}
     &(1.0,0.2)&Ours&(0.908,0.832)  &(1.035,0.893)  &(0.628,0.506) \\
     &         &    &(-0.885,-0.891)&(-1.042,-0.933)&(0.554,0.620) \\ \cline{3-6}
     &         &Ma's&(0.664,0.942)  &(0.715,1.033)  &(1.155,1.674) \\
     &         &    &(-1.352,-0.998)&(-0.801,-0.833)&(2.003,2.446) \\ \hline
  200&(3.0,1.0)&Ours&(0.694,0.502)  &(0.977,0.949)  &(0.788,0.831) \\
     &         &    &(-0.541,-0.497)&(-1.009,-0.959)&(0.861,0.855) \\ \cline{3-6}
     &         &Ma's&(0.984,1.254)  &(1.034,0.944)  &(1.100,1.198) \\
     &         &    &(-0.641,-0.946)&(-0.944,-0.965)&(0.918,0.677) \\ \cline{2-6}
     &(2.0,0.5)&Ours&(1.011,0.996)  &(1.012,0.988)  &(0.110,0.106) \\
     &         &    &(-1.004,-1.002)&(-1.005,-1.007)&(0.092,0.098) \\ \cline{3-6}
     &         &Ma's&(0.771,1.236)  &(0.982,0.966)  &(1.915,1.554) \\
     &         &    &(-0.642,-0.784)&(-0.917,-0.921)&(0.720,0.576) \\ \cline{2-6}
     &(1.0,0.2)&Ours&(0.901,0.948)  &(0.990,1.008)  &(0.451,0.442) \\
     &         &    &(-0.946,-0.912)&(-1.029,-0.977)&(0.478,0.420) \\ \cline{3-6}
     &         &Ma's&(0.655,0.717)  &(0.398,0.654)  &(1.584,1.281) \\
     &         &    &(-0.534,-1.306)&(-0.820,-0.933)&(1.622,1.187) \\
  \hline
\end{tabular}
\end{table}

\begin{table}[htbp]\small
\centering
\caption{Empirical mean, median and standard deviation of estimators of $(\theta^0_{11},\theta^0_{12})$ and $(\theta^0_{21},\theta^0_{22})$ for the \textit{case 2} in Example 2. \label{table5}}
\begin{tabular}{llllll}
  \hline
  $n$&$(\mu,\sigma)$&Methods&mean&median&std \\ \hline
  100&(3.0,1.0)&Ours&(1.637,1.708)  &(1.985,1.988)  &(1.185,0.959) \\
     &         &    &(-1.416,-1.481)&(-1.971,-1.935)&(1.266,1.196) \\ \cline{3-6}
     &         &Ma's&(1.893,1.599)  &(1.997,1.675)  &(0.824,0.914) \\
     &         &    &(-1.629,-1.983)&(-1.824,-2.004)&(0.832,0.595) \\ \cline{2-6}
     &(2.0,0.5)&Ours&(2.016,1.988)  &(2.018,1.994)  &(0.150,0.148) \\
     &         &    &(-1.937,-2.028)&(-1.977,-2.005)&(0.303,0.261) \\ \cline{3-6}
     &         &Ma's&(2.367,1.389)  &(2.246,1.590)  &(0.729,0.669) \\
     &         &    &(-1.243,-2.379)&(-1.388,-2.335)&(0.658,0.568) \\ \cline{2-6}
     &(1.0,0.2)&Ours&(1.915,1.942)  &(1.956,2.065)  &(0.591,0.596) \\
     &         &    &(-1.894-1.922) &(-1.923,-2.045)&(0.635,0.671) \\ \cline{3-6}
     &         &Ma's&(1.992,1.519)  &(2.737,2.291)  &(2.020,2.022) \\
     &         &    &(-2.750,-1.357)&(-2.405,-2.301)&(2.610,2.690) \\ \hline
  200&(3.0,1.0)&Ours&(1.962,2.003)  &(1.993,2.001)  &(0.111,0.057) \\
     &         &    &(-1.985,-2.005)&(-1.993,-2.005)&(0.085,0.043) \\ \cline{3-6}
     &         &Ma's&(1.910,1.693)  &(1.983,1.943)  &(1.345,1.892) \\
     &         &    &(-1.815,-1.803)&(-1.971,-1.956)&(0.851,0.743) \\ \cline{2-6}
     &(2.0,0.5)&Ours&(2.027,1.979)  &(2.030,1.972)  &(0.097,0.092) \\
     &         &    &(-1.981,-2.020)&(-1.978,-2.022)&(0.100,0.098) \\ \cline{3-6}
     &         &Ma's&(1.823,2.022)  &(1.946,1.891)  &(1.336,1.291) \\
     &         &    &(-1.946,-1.921)&(-2.006,-1.912)&(0.214,0.216) \\ \cline{2-6}
     &(1.0,0.2)&Ours&(1.936,2.057)  &(1.951,2.038)  &(0.227,0.235) \\
     &         &    &(-2.047,-1.958)&(-2.009,-1.940)&(0.269,0.252) \\ \cline{3-6}
     &         &Ma's&(1.523,2.741)  &(1.922,1.876)  &(1.033,1.821) \\
     &         &    &(-1.119,-1.100)&(-1.794,-1.786)&(1.492,1.825) \\
  \hline
\end{tabular}
\end{table}

\clearpage

\textit{Example 3 (Non-sparsity model)}. In this experiment, we consider the following non-sparsity model:
\begin{equation*}
  Y_i = X_i^T\beta + Z_i^T\theta_i + \epsilon_i, i = 1,\cdots, n,
\end{equation*} where $X$ and $Z$ are 3-dimensional variables.
Here the non-sparsity means that most of the parameters $\theta_1,\cdots,\theta_n$ are different. In this case, the parameters $\theta_1,\cdots,\theta_n$ are inestimable. However, by Theorem 3.1, our estimator $\widehat\beta$ defined in (\ref{(2.3)}) is a consistent estimator of $\beta$. In the following, we will illustrate this point of view. In the simulation procedure, $\beta$ is set as $\beta=(2,-2,3)^T$, and for non-sparsity, $\theta_1,\cdots,\theta_n$ are generated from 3-dimensional normal distribution $N(3\bm{1}, 4\bm{I})$. For a comprehensive comparison, we consider the following cases: \begin{itemize}
  \item[(1)] $X \sim N(\bm{\mu}_X, \Sigma_X)$, where $\bm{\mu}_X=(0,0,0)^T$ or $(2,2,2)^T$ or $(4,4,4)^T$, and $\Sigma_X = (\sigma_{ij})$ with $\sigma_{ij}=0.8^{|i-j|}$;
  \item[(1)] $Z \sim N(\bm{\mu}_Z, \Sigma_Z)$, where $\bm{\mu}_Z=(0,0,0)^T$ or $(1,1,1)^T$ or $(2,2,2)^T$, and $\Sigma_Z = (\sigma_{ij})$ with $\sigma_{ij}= 0.8^{|i-j|}$.
\end{itemize}
Furthermore, the correlation coefficient betweens each components of $X$ and $Z$ are the same as $\rho$. We compare our method with the ordinary least squares estimator $\widehat\beta_{LS}$ that ignores the inestimable parts $\theta_i$. The simulation results are listed in Table \ref{table6}. We have the following findings.

(1) It can be proved that when $Z$ has zero expectation and is uncorrelated with $X$, the OLS estimator is consistent theoretically. Even in this case, our estimator is much better than the OLS estimator in the sense that ours has smaller MSE and std.

(2) For all the cases, our estimator is always consistent and is much better than the OLS estimator.

(3) Our estimator is robust to the correlation between $X$ and $Z$, while the OLS estimator is very sensitive to the correlation.

\begin{table}[htbp]\small
\centering
\caption{Mean squares error (MSE) and std in Example 3. \label{table6}}
\begin{tabular}{lllllllllllll}
\hline
 \multicolumn{3}{c}{ } & & \multicolumn{2}{c}{$\bm{\mu}_X$}\\  \cline{4-7}
  \multicolumn{3}{c}{ } & &(0~0~0)  & (2~2~2) & (4~4~4)  \\ \hline
   $n$    & $\rho$  & $\bm{\mu}_Z$ &Methods&  MSE(std)  & MSE(std) & MSE(std) \\ \hline
   200  & 0.0 & (0~0~0) & OLS & 6.122(6.244) & 6.110(6.623) & 6.637(7.150)  \\
        &     &         & Ours & 0.955(1.060) & 0.7975(0.895) & 0.762(0.804)  \\
        &     & (1~1~1) & OLS & 15.218(15.864) & 17.974(11.377) & 10.461(8.827)  \\
        &     &         & Ours & 1.635(1.747)  & 1.732(1.785) & 1.7656(2.070) \\
        &     & (2~2~2) & OLS & 42.432(44.917) & 54.454(24.083) & 28.889(18.052)  \\
        &     &         & Ours & 4.586(4.872) & 3.812(4.196) & 4.172(4.577) \\   \cline{2-7}
        & 0.6 & (0~0~0) & OLS & 37.338(10.109) & 6.933(6.974) & 5.958(6.500)  \\
        &     &         & Ours& 0.881(0.870) & 0.8944(0.971) & 0.831(0.855) \\
        &     & (1~1~1) & OLS & 45.106(22.523) & 21.106(8.076) & 9.714(7.028)  \\
        &     &         & Ours & 1.531(1.611) & 1.6042(1.673) & 1.5843(1.687) \\
        &     & (2~2~2) & OLS & 68.494(48.883) & 61.032(16.935) & 21.779(10.198)  \\
        &     &         & Ours & 3.896(4.266) & 3.986(4.062) & 4.360(4.635) \\   \hline
   800  & 0.0 & (0~0~0) & OLS & 1.605(1.665) & 1.499(1.699) & 1.497(1.695)  \\
        &     &         & New & 0.221(0.228) & 0.361(0.403) & 0.411(0.450) \\
        &     & (1~1~1) & OLS & 3.581(3.722) & 12.938(4.369) & 5.277(2.688)  \\
        &     &         & Ours & 0.384(0.427) & 0.405(0.420) & 0.432(0.470) \\
        &     & (2~2~2) & OLS & 9.477(9.240) & 46.055(10.233) & 16.112(5.436)  \\
        &     &         & New & 1.008(1.093) & 0.924(0.942) & 1.0143(1.045) \\   \cline{2-7}
        & 0.6 & (0~0~0) & OLS & 32.790(4.451) & 2.384(1.781) & 1.524(1.520)  \\
        &     &         & Ours & 0.228(0.226) & 0.409(0.458) & 1.113(1.172) \\
        &     & (1~1~1) & OLS & 35.458(9.137) & 18.794(3.593) & 5.970(2.440)  \\
        &     &         & Ours & 1.025(1.051) & 0.971(1.041) & 1.023(1.087) \\
        &     & (2~2~2) & OLS & 40.152(18.036) & 58.266(7.749) & 17.433(4.224)  \\
        &     &         & Ours & 1.027(1.031) & 0.922(1.007) & 1.043(1.029) \\
        \hline
\end{tabular}
\end{table}

\clearpage

\subsection{Real data analysis}

In this section, we use real data to demonstrate the effectiveness of our method. The car sales data of USA are used to judge whether a potential consumer will buy an American car or a Japanese car according to his/her several characteristics. The data can be accessed at http://www-stat.wharton.upenn.edu/~waterman/fsw/baur/assigtxt.htm.
The dataset contains 259 consumption records, and each record consists of a binary response $Y$ and the covariate vector $X$ that has six components $X^{(j)},j=1,\cdots,6$, where
\begin{itemize} \item
$Y$ is the categorical variable with categories 0 (the Japanese car) and 1 (the American car);
\item $X^{(1)}$ is the consumer's age from 18 to 60;
\item $X^{(2)}$ is the consumer's gender with categories 0 (female) and 1 (male);
\item $X^{(3)}$ is the binary covariate: ``if consumer's age is 25'', with values 0 (No) and 1 (Yes);
\item $X^{(4)}$ is the consumer's marital status with categories 0 (single) and 1 (married);
\item $X^{(5)}$ is the favourable size of a consumer, the possible
levels being 0-2 (2 largest);
\item $X^{(6)}$ is the type of car with categories 0 (Work), 1 (Sporty) and 2 (Family);

\end{itemize}
Without loss of generality, we standardize each component in $X$ so that it has mean 0 and variance 1. For the response, we use the following transformation:
\begin{equation*}
  \widetilde{Y} = \log(\frac{a+Y}{b-Y}),
\end{equation*}
where $a=0.01, b=1.01$.

We first use a homogenous linear regression $\widetilde Y=X^T\beta+\epsilon$ to fit this dataset. By  least square estimate, we get the estimator $\widehat\beta$ of $\beta$, and the corresponding residual sum of squares as RSS$=\sum_{i=1}^{259}(\widetilde{Y}_i-\widehat{\widetilde{Y}}_i)^2=15.62$, in which $\widehat{\widetilde{Y}}_i = X_i^T \widehat{\beta}$, the fitted value of $\widetilde Y_i$. Such a relatively large value of RSS implies that the homogenous linear fitting may be unreasonable.
Therefore, a natural treatment is to examine if a heterogenous linear regression can fit the dataset better.
To this end, we first try the following partially heterogenous linear models with single heterogenous coefficient as
\begin{equation}\label{try_model} \mbox{Model } l:
  \widetilde{Y}_i = {{\beta}^{(-l)}}^T X_i^{(-l)} + \theta_{i}^{(l)}X_{i}^{(l)} + \epsilon_i, \ i=1,\cdots,259,\ l=1,\cdots,6,
\end{equation}
where $X_i^{(-l)}$ is the covariate vector without the $l$-th component $X_{i}^{(l)}$ of $X_{i}$.
For $l=1,\cdots,6$, the corresponding RSS's of the six models in (\ref{try_model})  are reported in table \ref{rss_6}.
\captionsetup[table]{justification=centering,singlelinecheck=off,font=small}
\begin{table}[htbp]
\centering
\caption{RSS of the six models in (\ref{try_model})}\label{rss_6}\small
\begin{tabular}{c|cccccc}
  \hline
  Models &Model 1&Model 2&Model 3&Model 4&Model 5&Model 6  \\
  RSS                   &12.52&15.56 &3.91 &2.83 &3.27 &6.40   \\
  \hline
\end{tabular}
\end{table}
From Table \ref{rss_6}, we see that Model 4 and Model 5 that respectively contain heterogenous coefficients $\theta_i^{(4)}$ and $\theta_i^{(5)}$ can reduce the RSS obviously. Then, we combine the covariates $X^{(4)}$ and $X^{(5)}$ together as the heterogenous part to reconstruct the following heterogenous model:
\begin{equation}\label{try_model2}
  \widetilde{Y}_i = \beta^{(1)}X_{i}^{(1)}+\beta^{(2)}X_{i}^{(2)}+\beta^{(3)}X_{i}^{(3)}+\beta^{(6)}X_{i}^{(6)}
  +\theta_{i}^{(4)}X_{i}^{(4)}+\theta_{i}^{(5)}X_{i}^{(5)} + \epsilon_i.
\end{equation}
The RSS of model (\ref{try_model2}) is significantly reduced to 1.13. It ensures that the partially heterogenous linear model with the two heterogenous coefficients can significantly improve the fitting.

It is worth noting that it seems that $X^{(3)}$ may be another heterogenous variable as the corresponding heterogenous Model 3 has the third-smallest RSS. However, if we use the three variables $X^{(3)}$, $X^{(4)}$ and $X^{(5)}$ together as the heterogenous part, the resulting RSS is 6.49, larger than that obtained by the model with the two heterogenous variables. Therefore, model (\ref{try_model2}) is our final choice.

By the above model, the 259 data are classified into 4 subgroups and  the subgroups respectively contain 108, 74, 41 and 36 data. From the classification analysis, we see that about 62\% people who prefer the large size car choose the American car, while about 78\% people who prefer the small car choose the Japanese car.
On the other hand, it is found that the married people are more inclined to buy American car, as among the American car owners, about 72\% people are married, while among the Japanese car owners, only 58\% people are married. These facts are consistent with the empirical analysis, as we all know, Japanese cars are usually smaller than the American ones. In short, the people who either prefer the large size car or are married tend to buy the American car, otherwise the people are like to buy the Japanese car. This provides the car dealers with useful information to promote their cars.

\

\leftline{\Large\bf References}

\begin{description}

\item Banfield, J. D. and Raftery, A. E. (1993). Model-based gaussian and non-gaussian clustering.
{\it Biometrics}, {\bf 49}, 803-821.

\item Bondell, H. D. and Reich, B. J. (2008). Simultaneous regression shrinkage, variable selection, and
supervised clustering of predictors with oscar. {\it Biometrics}, {\bf 64}, 115-123.

\item Boyd, S., Parikh, N., Chu, E., Peleato, B., and Eckstein, J. (2011). Distributed optimization and
statistical learning via the alternating direction method of multipliers. {\it Foundations and Trends
in Machine Learning}, {\bf 3}, 1-122.

\item Everitt, B. and Hand, D. J. (1981). {\it Finite Mixture Distributions}. Chapman and Hall, New York.

\item  Chi, E. C. and Lange, K. (2014). Splitting methods for convex clustering. {\it Journal of Computational and Graphical Statistics}, {\bf 24}, 994-1013.

\item Guo, F. J., Levina, E., Michailidis, G. and Zhu, J. (2010). Pairwise variable selection for high dimensional
model-based clustering. {\it Biometrics}, {\bf 66}, 793-804.

\item Guo, J., Hu, J., Jing, B. and Zhang, Z. (2016). Spline-lasso in high-dimensional linear
regression. {\it J Amer Statist Assoc}, (to appear).

\item Gordon, R. D. (1941). Values of mills'ratio of area to bounding ordinate and of the nor-
mal probability integral for large values of the argument. {\it The Annals of Mathematical
Statistics}, {\bf 12}, 364-366.

\item Hastie, T. and Tibshirani, R. (1996). Discriminant analysis by gaussian mixtures. {\it Journal of the
Royal Statistical Society}, Series B, {\bf 58}, 155-176.


\item Ke, T., Fan, J. and Wu, Y. (2015). Homogeneity in regression. {\it Journal of the American Statistical
Association}, {\bf 110}, 175-194.

\item Osipov, L. V. and Petrov, V. V. (1967). Short communications on an estimate of the remainder term in the central limit theorem. {\it Theory of Probability and Its applications}, {\bf 12}, 281-286.

\item Shen, X. and Huang, H. C. (2010). Grouping pursuit through a regularization solution surface.
{\it Journal of the American Statistical Association}, {\bf 105}, 727-739.

\item  Lin, L. , Dong, P. , Song, Y.  and Zhu, L.  (2016).  Upper expectation parametric regression.  {\it Statistica Sinica}, {\bf 27}, 1265-1280.

\item Ma, S. and Huang, J. (2016a). A concave pairwise fusion approach to subgroup
analysis. {\it J Amer Statist Assoc.} http://www.tandfonline.com/doi/abs/10.1080/ 01621459.2016.1148039.

\item Ma, S. and Huang, J. (2016b). Estimating subgroup-specific treatment
effects via concave fusion. (Manuscript)

\item McNicholas, P. D. (2010). Model-based classification using latent gaussian mixture models. {\it Journal of Statistical Planning and Inference}, {\bf 140}, 1175-1181.

\item Shen, J. and He, X. (2015). Inference for subgroup analysis with a structured logistic-normal
mixture model. {\it J Amer Statist Assoc}, {\bf 110}, 303-312.

\item Tibshirani, R. and Saunders, M. (2005). Sparsity and smoothness via the fused lasso.
{\it J. R. Statist. Soc}. B. {\bf 67}, 91-108.

\item You, Z.,  Si, Y. W., Zhang, D., Zeng, X. X., Leung S. C . H. and Li, T. (2015). A decision-making framework for precision marketing. {\it Expert Systems with Applications}, {\bf 42}, 3357-3367.

\item Zabin, J. and Brebach, G. (2004). {\it Precision Marketing: The New Rules for Attracting, Retaining, and Leveraging Profitable Customers.}  John Wiley \& Sons, Inc.

\end{description}

\setcounter{equation}{0}
\section{Appendix: Proofs}

\noindent{\it Proof of Theorem 3.1.} In this proof, the true values are still denoted by $\beta$ and  $\overline\theta_{\mathscr A}$ for the simplicity of the notation. We first prove the first result.
It follows from model (\ref{(1.1)}) that
\begin{eqnarray*}\overline{XY}&=&\overline{XX^T}\beta+\frac{1}{n}\sum_{i=1}^nX_iZ_i^T\theta_i
+\frac{1}{n}\sum_{i=1}^nX_i\epsilon_i,\\
\overline{ZY}&=&\overline{ZX^T}\beta+\frac{1}{n}\sum_{i=1}^nZ_iZ_i^T\theta_i
+\frac{1}{n}\sum_{i=1}^nZ_i\epsilon_i.\end{eqnarray*}
Then
\begin{eqnarray*}&&\overline{XY}-\overline{XZ^T}(\overline{ZZ^T})^{-1}\overline{ZY}\\&&=\left(\overline{X X^T}-\overline{XZ^T}(\overline{ZZ^T})^{-1}\overline{Z X^T}\right)\beta+\frac{1}{n}\sum_{i=1}^nX_iZ_i^T\left(\theta_i-(\overline{ZZ^T})^{-1}
\frac{1}{n}\sum_{i=1}^nZ_iZ_i^T\theta_i\right)
\\&& \ \ \ \
+\frac{1}{n}\sum_{i=1}^n\left(X_i-\overline{XZ^T}(\overline{ZZ^T})^{-1}Z_i\right)\epsilon_i.
\end{eqnarray*}
By the result above and the estimator (\ref{(2.3)}), we have
\begin{eqnarray*}\widehat\beta-\beta&=&\left(\overline{X X^T}-\overline{XZ^T}(\overline{ZZ^T})^{-1}\overline{Z X^T}\right)^{-1}\frac{1}{n}\sum_{i=1}^nX_iZ_i^T\left(\theta_i-(\overline{ZZ^T})^{-1}
\frac{1}{n}\sum_{i=1}^nZ_iZ_i^T\theta_i\right)
\\&&  +\left(\overline{X X^T}-\overline{XZ^T}(\overline{ZZ^T})^{-1}\overline{Z X^T}\right)^{-1}\frac{1}{n}\sum_{i=1}^n\left(X_i-\overline{XZ^T}(\overline{ZZ^T})^{-1}Z_i\right)\epsilon_i\end{eqnarray*}
Note that $$\overline{X X^T}-\overline{XZ^T}(\overline{ZZ^T})^{-1}\overline{Z X^T}\stackrel{P}\longrightarrow \Omega=E[X X^T]-E[XZ^T](E[ZZ^T])^{-1}E[Z X^T].$$ Then $\sqrt n\,\Omega\,(\widehat\beta-\beta)$ is asymptotically identically distributed as
\begin{eqnarray}\label{a1}\frac{1}{\sqrt n}\sum_{i=1}^nX_iZ_i^TW_i +\frac{1}{\sqrt n}\sum_{i=1}^n\left(X_i-\overline{XZ^T}(\overline{ZZ^T})^{-1}Z_i\right)\epsilon_i,\end{eqnarray}
where the weight $W_i=\left(\theta_i-(\overline{ZZ^T})^{-1}
\frac{1}{n}\sum_{i=1}^nZ_iZ_i^T\theta_i\right)$. By the law of large numbers, we have $W_i\stackrel{P}\longrightarrow\left(\theta_i-
\overline\theta\right)$. Thus, the weighted sum $\frac{1}{\sqrt n}\sum_{i=1}^nX_iZ_i^TW_i$ is equal to $$\frac{1}{\sqrt n}\sum_{i=1}^nX_iZ_i^T\left(\theta_i-
\overline\theta\right)+o_p(1)\frac{1}{\sqrt n}\sum_{i=1}^nX_iZ_i^T\left(\theta_i-
\overline\theta\right)$$ and then is asymptotically identically distributed as $\frac{1}{\sqrt n}\sum_{i=1}^nX_iZ_i^T\left(\theta_i-
\overline\theta\right)$. It can be seen that
\begin{eqnarray*}E\left[\frac{1}{\sqrt n}\sum_{i=1}^nX_iZ_i^T\left(\theta_i-
\overline\theta\right)\right]=0\end{eqnarray*}
and
\begin{eqnarray*}&& Var\left[\frac{1}{\sqrt n}\sum_{i=1}^nX_iZ_i^T\left(\theta_i-
\overline\theta\right)\right]\\&&=\frac{1}{n}\sum_{i=1}^n\left(\theta_i-
\overline\theta\right)^TE\{(Z X^T-E[Z X^T])(Z X^T-E[Z X^T])^T\}\left(\theta_i-
\overline\theta\right).\end{eqnarray*}
Therefore,
\begin{eqnarray}\label{a2}\frac{1}{\sqrt n}\sum_{i=1}^nX_iZ_i^T\left(\theta_i-
\overline\theta\right)\stackrel{D}\longrightarrow N(0,\Psi).\end{eqnarray} Moreover, it is clear that
\begin{eqnarray}\label{a3}\frac{1}{\sqrt n}\sum_{i=1}^n\left(X_i-\overline{XZ^T}(\overline{ZZ^T})^{-1}Z_i\right)\epsilon_i\stackrel{D}\longrightarrow N(0,\Phi).\end{eqnarray} By combining (\ref{a1}), (\ref{a2}), (\ref{a3}) and the uncorrelation between $(X,Z)$ and $\epsilon$, we can prove the first result of the theorem.

In the following, we prove the second result. By model (\ref{(1.1)}), the estimator (\ref{(2.4)}) can be expressed as
\begin{eqnarray*}&&\hspace{-0.8cm}\widehat{\overline\theta}_{\mathscr A}=(\overline{ZZ^T}_{\mathscr A})^{-1}\left(\overline{ZY}_{\mathscr A}-\overline{Z X^T}_{\mathscr A}\widehat\beta\right)\\&=&(\overline{ZZ^T}_{\mathscr A})^{-1}\left(\overline{ZX^T}_{\mathscr A}\beta+\frac{1}{|\mathscr A|}\sum_{i\in\mathscr A}Z_iZ_i^T\theta_i
+\frac{1}{|\mathscr A|}\sum_{i\in\mathscr A}Z_i\epsilon_i-\overline{Z X^T}_\mathscr A\widehat\beta\right)\\&=&
\overline\theta_{\mathscr A}+(\overline{ZZ^T}_{\mathscr A})^{-1}\left(\frac{1}{|\mathscr A|}\sum_{i\in\mathscr A}Z_iZ_i^T\left(\theta_i-\overline\theta\right)
+\frac{1}{|\mathscr A|}\sum_{i\in\mathscr A}Z_i\epsilon_i-\overline{Z X^T}_\mathscr A(\widehat\beta-\beta)\right).\end{eqnarray*} As shown above,
$\sqrt n\,\Omega\,(\widehat\beta-\beta)$ is asymptotically identically distributed as
\begin{eqnarray*}\frac{1}{\sqrt n}\sum_{i=1}^nX_iZ_i^T\left(\theta_i-\overline\theta\right) +\frac{1}{\sqrt n}\sum_{i=1}^n\left(X_i-\overline{XZ^T}(\overline{ZZ^T})^{-1}Z_i\right)\epsilon_i.\end{eqnarray*}
The above results imply that $\sqrt n\left(\widehat{\overline\theta}_\mathscr A-\overline\theta_\mathscr A\right)$ is asymptotically identically distributed as
\begin{eqnarray*}&&(\overline{ZZ^T}_\mathscr A)^{-1}\frac{1}{\sqrt {|\mathscr A|}}\sum_{i\in\mathscr A}\left(Z_iZ_i^T-\overline{Z X^T}\Omega^{-1}X_iZ_i^T\right)\left(\theta_i-\overline\theta\right)\\&&+
(\overline{ZZ^T})^{-1}\left(\frac{1}{\sqrt {|\mathscr A|}}\sum_{i\in\mathscr A}
\left(Z_i-\Omega^{-1}\left(X_i-\overline{XZ^T}(\overline{ZZ^T})^{-1}Z_i\right)\right)\epsilon_i\right).\end{eqnarray*}
It can be easily verified that the above is asymptotically distributed as following normal distribution: $$N\left(0,(E[ZZ^T])^{-1}(\Lambda_\mathscr A+\sigma^2\Upsilon)(E[ZZ^T])^{-1}\right).$$ $\Box$

\

\noindent{\it Proof of Lemma 3.2.} In this proof, the true values are still denoted by $\beta$ and  $\alpha_j$ for the simplicity of the notation.
By the definition, we get
\begin{eqnarray*}\nonumber&&\left((\widehat\alpha_1^0)^T,\cdots,(\widehat\alpha_k^0)^T\right)^T
\\&&\nonumber=({\mathscr I}^T{\mathscr I})^{-1}{\mathscr I}^T{\bf Z}\circ({\bf Y}-
{\bf X}\widehat\beta)\\&&\nonumber=\left(m_1^{-1}\sum_{i\in\mathscr G_1}(Y_i-X_i^T\widehat\beta)Z_i^T,\cdots,m_k^{-1}\sum_{i\in\mathscr G_k}(Y_i-X_i^T\widehat\beta)Z_i^T\right)^T\\&&\nonumber=\left(m_1^{-1}\sum_{i\in\mathscr G_1}(\theta_i^T Z_i+(\beta-\widehat\beta)^T X_i)Z_i^T,\cdots,m_k^{-1}\sum_{i\in\mathscr G_k}(\theta_i^T Z_i+(\beta-\widehat\beta)^T X_i)Z_i^T\right)^T\\&&\nonumber \ \ \ +\left(m_1^{-1}\sum_{i\in\mathscr G_1}Z_i^T\epsilon_i,\cdots,m_k^{-1}\sum_{i\in\mathscr G_k}Z_i^T\epsilon_i\right)^T\\&&\nonumber=\left(m_1^{-1}\sum_{i\in\mathscr G_1}(\alpha_i^T+(\beta-\widehat\beta)^TX_iZ_i^T),\cdots,m_k^{-1}\sum_{i\in\mathscr G_k}(\alpha_i^T+(\beta-\widehat\beta)^TX_iZ_i^T)\right)^T\\&&\nonumber \ \ \ +\left(m_1^{-1}\sum_{i\in\mathscr G_1}Z_i^T\epsilon_i,\cdots,m_k^{-1}\sum_{i\in\mathscr G_k}Z_i^T\epsilon_i\right)^T\\&&\ \ \ +\left(m_1^{-1}\sum_{i\in\mathscr G_1}(\theta^TZ_iZ_i^T-\alpha_i),\cdots,m_k^{-1}\sum_{i\in\mathscr G_k}(\theta^TZ_iZ_i^T-\alpha_i)\right)^T\\&&\nonumber=\left(\alpha_1^T+m_1^{-1}\sum_{i\in\mathscr G_1}(\beta-\widehat\beta)^TX_iZ_i^T,\cdots,\alpha_k^T+m_k^{-1}\sum_{i\in\mathscr G_k}(\beta-\widehat\beta)^TX_iZ_i^T\right)^T\\&&\nonumber \ \ \ +\left(m_1^{-1}\sum_{i\in\mathscr G_1}Z_i^T\epsilon_i,\cdots,m_k^{-1}\sum_{i\in\mathscr G_k}Z_i^T\epsilon_i\right)^T\\&&\ \ \ +\left(m_1^{-1}\sum_{i\in\mathscr G_1}(\theta^TZ_iZ_i^T-\alpha_i),\cdots,m_k^{-1}\sum_{i\in\mathscr G_k}(\theta^TZ_iZ_i^T-\alpha_i)\right)^T.\end{eqnarray*} Then,
\begin{eqnarray*}\nonumber&&\left(\sqrt{m_1}(\widehat\alpha_1^0-\alpha_1)^T,\cdots,
\sqrt{m_k}(\widehat\alpha_k^0-\alpha_k)^T\right)^T
\\&&\nonumber=\left(\sqrt{m_1}^{-1}\sum_{i\in\mathscr G_1}X_i Z_i^T,\cdots,\sqrt{m_k}^{-1}\sum_{i\in\mathscr G_k}X_i Z_i^T\right)^T(\beta-\widehat\beta)\\&&\nonumber \ \ \ +\left(\sqrt{m_1}^{-1}\sum_{i\in\mathscr G_1}Z_i\epsilon_i,\cdots,\sqrt{m_k}^{-1}\sum_{i\in\mathscr G_k}Z_i\epsilon_i\right)^T\\&&\ \ \ +\left(\sqrt{m_1}^{-1}\sum_{i\in\mathscr G_1}(\theta^TZ_iZ_i^T-\alpha_i),\cdots,\sqrt{m_k}^{-1}\sum_{i\in\mathscr G_k}(\theta^TZ_iZ_i^T-\alpha_i)\right)^T.\end{eqnarray*}
The proof of Theorem 3.1 shows that
$\sqrt n\,\Omega\,(\widehat\beta-\beta)$ is asymptotically identically distributed as
\begin{eqnarray*}\frac{1}{\sqrt n}\sum_{i=1}^nX_iZ_i^T\left(\theta_i-\overline\theta\right) +\frac{1}{\sqrt n}\sum_{i=1}^n\left(X_i-\overline{XZ^T}(\overline{ZZ^T})^{-1}Z_i\right)\epsilon_i.\end{eqnarray*}
Thus, $\left((\widehat\alpha_1^0)^T,\cdots,(\widehat\alpha_k^0)^T\right)^T$ is asymptotically identically distributed as
\begin{eqnarray*}&&\left(\sqrt{m_1}^{-1}\sum_{i\in\mathscr G_1}X_i Z_i^T,\cdots,\sqrt{m_k}^{-1}\sum_{i\in\mathscr G_k}X_i Z_i^T\right)^T\\&&\ \ \ \times\left(\frac{1}{n}\Omega^{-1}\sum_{i=1}^nX_iZ_i^T\left(\theta_i-\overline\theta\right) +\frac{1}{\sqrt n}\sum_{i=1}^n\left(X_i-\overline{XZ^T}(\overline{ZZ^T})^{-1}Z_i\right)\epsilon_i\right)\\&&\nonumber \ \ \ +\left(\sqrt{m_1}^{-1}\sum_{i\in\mathscr G_1}Z_i\epsilon_i,\cdots,\sqrt{m_k}^{-1}\sum_{i\in\mathscr G_k}Z_i\epsilon_i\right)^T\\&&\ \ \ +\left(\sqrt{m_1}^{-1}\sum_{i\in\mathscr G_1}(\theta^TZ_iZ_i^T-\alpha_i),\cdots,\sqrt{m_k}^{-1}\sum_{i\in\mathscr G_k}(\theta^TZ_iZ_i^T-\alpha_i)\right)^T.\end{eqnarray*}
By the above result, the independence between $(X_i^T,Z_i^T)$ and $\epsilon_i$, together with the Central Limit
Theorem, we can prove the theorem. $\Box$

\

\noindent{\it Proof of Lemma 3.3.} Let
$${\bf Q}=\left({\bf X}^T{\bf X}-{\bf X}^T {\bf Z}({\bf Z}^T{\bf Z})^{-1}{\bf Z}^T {\bf X}\right)^{-1}\left({\bf X}^T-{\bf X}^T{\bf Z}({\bf Z}^T{\bf Z})^{-1}{\bf Z}^T\right).$$
By (\ref{(2.3)}), we have
$\widehat\beta-\beta={\bf Q}{\bbf\epsilon}.$  It follows from the proof of Lemma 3.2 that
\begin{eqnarray*}&&((\widehat\alpha_1^0-\alpha_1)^T,\cdots,
(\widehat\alpha_k^0-\alpha_k)^T)^T
\\&&=\left(m_1^{-1}\sum_{i\in\mathscr G_1}X_i Z_i^T,\cdots,m_k^{-1}\sum_{i\in\mathscr G_k}X_i Z_i^T\right)^T{\bf Q}{\bbf\epsilon}\\&&\ \ \ +\left(m_1^{-1}\sum_{i\in\mathscr G_1}Z_i^T\epsilon_i,\cdots,m_k^{-1}\sum_{i\in\mathscr G_k}Z_i^T\epsilon_i\right)^T \\&&\ \ \ + \left(m_1^{-1}\sum_{i\in\mathscr G_1}\theta_i^T (Z_iZ_i^T-E[Z_iZ_i^T]),\cdots,m_k^{-1}\sum_{i\in\mathscr G_k}\theta_i^T (Z_i Z_i^T-E[Z_iZ_i^T])\right)^T\\&&=\left(m_1^{-1}\sum_{i\in\mathscr G_1}X_i Z_i^T,\cdots,m_k^{-1}\sum_{i\in\mathscr G_k}X_i Z_i^T\right)^TE[{\bf Q}]{\bbf\epsilon}\\&&\ \ \ +\left(m_1^{-1}\sum_{i\in\mathscr G_1}X_i Z_i^T,\cdots,m_k^{-1}\sum_{i\in\mathscr G_k}X_i Z_i^T\right)^T({\bf Q}-E[{\bf Q}]){\bbf\epsilon}\\&&\ \ \ +\left(m_1^{-1}\sum_{i\in\mathscr G_1}Z_i^T\epsilon_i,\cdots,m_k^{-1}\sum_{i\in\mathscr G_k}Z_i^T\epsilon_i\right)^T \\&&\ \ \ + \left(m_1^{-1}\sum_{i\in\mathscr G_1}\theta_i^T (Z_iZ_i^T-E[Z_iZ_i^T]),\cdots,m_k^{-1}\sum_{i\in\mathscr G_k}\theta_i^T (Z_i Z_i^T-E[Z_iZ_i^T])\right)^T.\end{eqnarray*} Note that
\begin{eqnarray*}&&\left\|m_1^{-1}\sum_{i\in\mathscr G_1}\theta_i^T (Z_iZ_i^T-E[Z_iZ_i^T]),\cdots,m_k^{-1}\sum_{i\in\mathscr G_k}\theta_i^T (Z_i Z_i^T-E[Z_iZ_i^T])\right\|_\infty \\&&
\leq |\mathscr G|_{\min}^{-1}\left\|\sum_{i\in\mathscr G_1}\theta_i^T (Z_iZ_i^T-E[Z_iZ_i^T]),\cdots,\sum_{i\in\mathscr G_k}\theta_i^T (Z_i Z_i^T-E[Z_iZ_i^T])\right\|_\infty,\end{eqnarray*}
and
\begin{eqnarray*}\left\|m_1^{-1}\sum_{i\in\mathscr G_1}Z_i^T\epsilon_i,\cdots,m_k^{-1}\sum_{i\in\mathscr G_k}Z_i^T\epsilon_i\right\|_\infty
\leq |\mathscr G|_{\min}^{-1}\left\|\sum_{i\in\mathscr G_1}Z_i^T\epsilon_i,\cdots,\sum_{i\in\mathscr G_k}Z_i^T\epsilon_i\right\|_\infty.\end{eqnarray*}
Let $Z^{(l)}_i$ be the $l$-th element of $Z_i$.
By the conditions {\it C4} and {\it C6}, we have that
\begin{eqnarray*}&&
P\left(\left\|\sum_{i\in\mathscr G_1}Z_i^T\epsilon_i,\cdots,\sum_{i\in\mathscr G_k}Z_i^T\epsilon_i\right\|_\infty>\sqrt{n\log n}\right)\\&&\leq \sum_{j=1}^k\sum_{l=1}^{d_Z}P\left(\Big|\sum_{i\in\mathscr G_j}Z_i^{(l)}\epsilon_i\Big|>\sqrt{n\log n}\right)\\&&\leq
2kd_Z\exp\left(-c_2\log n\right)=2kd_Zn^{-c_2}.
\end{eqnarray*} Let $\theta_i^{(l)}$ be the $l$-th element of $\theta_i$.
By the same argument as used above, we have
\begin{eqnarray*}&&P\left(\left\|\sum_{i\in\mathscr G_1}\theta_i^T (Z_iZ_i^T-E[Z_iZ_i^T]),\cdots,\sum_{i\in\mathscr G_k}\theta_i^T (Z_i Z_i^T-E[Z_iZ_i^T])\right\|_\infty>\sqrt{n\log n}\right)\\&&
\leq\sum_{1\leq l, s\leq d_Z}P\left(\Big|\sqrt{m_j}^{-1}\sum_{i\in\mathscr G_j}\left(\theta^{(l)}Z_i^{(l)}Z_i^{(s)}-E[\theta^{(l)}Z_i^{(l)}Z_i^{(s)}]\right)\Big|>\sqrt{\log n}\right)
\\&&=\sum_{1\leq l, s\leq d_Z}2\left[1-\Phi\left(\sqrt{\log n}\right)+\left(\Phi\left(\sqrt{\log n}\right)-F_{m_j}\left(\sqrt{\log n}\right)\right)\right]\\&&\leq\sum_{1\leq l, s\leq d_Z}2\left[1-\Phi\left(\sqrt{\log n}\right)+\left(\Phi\left(\sqrt{\log n}\right)-F_{|\mathscr G|_{\min}}\left(\sqrt{\log n}\right)\right)\right],
\end{eqnarray*} where $\Phi(u)$ is the distribution function of standard normal and $F_{m}(u)$ is the distribution function of
$\frac{1}{\sqrt{m}}\sum_{i=1}^m(Z_i^{(l)}Z_i^{(s)}-E[Z_i^{(l)}Z_i^{(s)}])$. By the theorem on the convergence rate in the central limit theorem (see, e.g., Osipov and Petrov 1967), we have $|\Phi(u)-F_{m}(u)|\leq \varphi(\sqrt m(1+|u|))m^{-\delta/2}u^{-2-\delta}$, where $\varphi(u)$ is a certain function, defined in the region $u>0$, bounded and non-increasing with $\lim\limits_{u\rightarrow\infty}\varphi(u)=0$. Moreover, by the inequality for the Gaussian tail probability (Gordon, 1941), we have
$1-\Phi(u)<u^{-1}e^{-\frac12 u^2}.$
Then,
\begin{eqnarray*}&&\left(\left\|\sum_{i\in\mathscr G_1}\theta_i^T (Z_iZ_i^T-E[Z_iZ_i^T]),\cdots,\sum_{i\in\mathscr G_k}\theta_i^T (Z_i Z_i^T-E[Z_iZ_i^T])\right\|_\infty>\sqrt{n\log n}\right)\\&&
\leq 2d_Xd_Z\sum_{j=1}^k\left[\left(\sqrt{\log n}\right)^{-1}\exp\left(-\frac12\log n\right)+\varphi(\sqrt{|\mathscr G|_{\min}}(1+\log n))|\mathscr G|_{\min}^{-\delta/2}\left(\sqrt{\log n}\right)^{-2-\delta}\right]\\&&=
2d_Xd_Z\left[\left(\sqrt{\log n}\right)^{-1}n^{-1/2}+\varphi(\sqrt{|\mathscr G|_{\min}}(1+\log n))|\mathscr G|_{\min}^{-\delta /2}\left(\sqrt{\log n}\right)^{-2-\delta}\right].
\end{eqnarray*}
Furthermore,
\begin{eqnarray*}&&
\left\|\left(m_1^{-1}\sum_{i\in\mathscr G_1}X_i Z_i^T,\cdots,m_k^{-1}\sum_{i\in\mathscr G_k}X_i Z_i^T\right)^TE[{\bf Q}]{\bbf\epsilon}\right\|_\infty\\&&\leq
\left\|\left(E[XZ^T],\cdots,E[XZ^T]\right)^TE[{\bf Q}]{\bbf\epsilon}\right\|_\infty\\&&\ \ \ +
\left\|\left(m_1^{-1}\sum_{i\in\mathscr G_1}(X_i Z_i^T-E[X_i Z_i^T]),\cdots,m_k^{-1}\sum_{i\in\mathscr G_k}(X_i Z_i^T-E[X_i Z_i^T])\right)^TE[{\bf Q}]{\bbf\epsilon}\right\|_\infty. \end{eqnarray*} Then, when $n$ is large enough,
\begin{eqnarray*}&&
P\left(\left\|\left(E[XZ^T],\cdots,E[XZ^T]\right)^TE[{\bf Q}]{\bbf\epsilon}\right\|_\infty>\sqrt{n\log n}\right)\\&&\leq\sum_{i=1}^{kd_Z}
P\left(\Big|\sum_{j=1}^n q_{ij}\epsilon_j\Big|>\sqrt{n\log n}\right)
\\&&\leq kd_Z
P\left(\Big|\sum_{j=1}^n q_{ij}\epsilon_j\Big|>\sqrt{n\log n}\right)\\&&\leq
2kd_Z\exp\left(-c_1c_2\log n\right)=2kd_Zn^{-c_1c_2}. \end{eqnarray*} Note that
$m_j^{-1}|\sum_{i\in\mathscr G_j}(X_i Z_i^T-E[X_i Z_i^T])|=0$ (a.s.) when $m_j$ goes to infinity. By the same argument as used above, we have that when $m_j$ is large enough,
\begin{eqnarray*}&&P\left(\left\|\left(m_1^{-1}\sum_{i\in\mathscr G_1}(X_i Z_i^T-E[X_i Z_i^T]),\cdots,m_k^{-1}\sum_{i\in\mathscr G_k}(X_i Z_i^T-E[X_i Z_i^T])\right)^TE[{\bf Q}]{\bbf\epsilon}\right\|_\infty>\sqrt{n\log n}\right)\\&&\leq
2kd_Zn^{-c_1c_2}.\end{eqnarray*}

Therefore, by combining the conclusions above, we have that with probability at least $1-\delta_n$, the following holds:
\begin{eqnarray*}\left\|((\widehat\alpha_1^0-\alpha_1)^T,\cdots,
(\widehat\alpha_k^0-\alpha_k)^T)^T\right\|_\infty\leq
|\mathscr G|_{\min}^{-1}\sqrt{n\log n}.\end{eqnarray*}  $\Box$

\noindent{\it Proof of Lemma 3.4.}
Define \begin{eqnarray*}&&L({\bbf \alpha})=\frac {1}{2}\sum_{i=1}^{n}\left(Z_i(Y_i-X_i^T\widehat\beta)-\alpha_i\right)^2,  P({\bbf \alpha})=\sum_{i<j}p_\gamma(|\alpha_i-\alpha_j|,\lambda),\\&&L_T({\bbf\alpha}^{0})
=\frac{1}{2}\sum_{j=1}^k\sum_{i\in\mathscr G_j}\left(Z_i(Y_i-X_i^T\widehat\beta)-\alpha_j^{0}\right)^2,  P_T({\bbf\alpha}^{0})=\sum_{i<j}m_im_j p_\gamma(|\alpha^{0}_{i}-\alpha^{0}_{j}|,\lambda),\end{eqnarray*}
and
$$Q({\bbf \alpha})=L_T({\bbf\alpha})+P({\bbf \alpha}),\ Q_T({\bbf\alpha}^{0})=L_T({\bbf\alpha}^{0})+P_T({\bbf\alpha}^{0}).$$
Then, the remainder of proof is the same as that of proving Theorem 2 of Ma and Huang (2016a).
 $\Box$

\

\noindent{\it Proof of Theorem 3.5.} It follows directly from Lemmas 3.2-3.4. $\Box$

\end{document}